\documentclass[10pt,onecolumn,journal,draftcls]{IEEEtran}

\usepackage{graphicx}
\usepackage{epsfig} 
\usepackage{amsmath} 
\usepackage{amssymb}  
\usepackage{mathrsfs}
\usepackage{url}
\usepackage{tabularx,ragged2e,booktabs,caption}
\usepackage{bm}
\usepackage{bbm}
\usepackage{color}
\usepackage{theorem}
\usepackage{caption}
\usepackage{subcaption}
\usepackage{supertabular}
\usepackage{hyperref}
\usepackage{float}
\usepackage{longtable}
\usepackage{array}
\usepackage{url}
\usepackage{booktabs}
\usepackage{multirow}
\usepackage{stfloats}

\allowdisplaybreaks

\makeatletter
\newcommand{\rmnum}[1]{\romannumeral #1}
\newcommand{\Rmnum}[1]{\expandafter\@slowromancap\romannumeral #1@}
\makeatother

\newtheorem{thm}{Theorem}
\newtheorem{lemma}{Lemma}

\newtheorem{rem}{Remark}

\newtheorem{ass}{Assumption}

\def\bx{{\mathbf x}}  

\def\bz{{\mathbf z}}

\def\bI{{\mathbf I}}

\def\bO{{\mathbf O}}
\def\bP{{\mathbf P}}
\def\bQ{{\mathbf Q}}

\def\bU{{\mathbf U}}

\def\bY{{\mathbf Y}}

\def\bZ{{\mathbf Z}}

\def\texitem#1{\par\smallskip\noindent\hangindent 25pt
               \hbox to 25pt {\hss #1 ~}\ignorespaces}

\newcommand{\bzero}{{\mathbf{0}}}

\newcommand{\scrA}{\mathcal{A}}

\newcommand{\scrJ}{\mathcal{J}}

\newcommand{\scrL}{\mathcal{L}}

\newcommand{\bGamma}{\boldsymbol{\Gamma}}

\newcommand{\bTheta}{\boldsymbol{\Theta}}

\newcommand{\blambda}{\boldsymbol{\lambda}}
\newcommand{\bLambda}{\boldsymbol{\Lambda}}
 
\newcommand{\bnu}{{\boldsymbol{\nu}}}
\newcommand{\bpi}{{\boldsymbol{\pi}}}

\newcommand{\bSigma}{{\boldsymbol{\Sigma}}}

\newcommand{\tsum}{\sum\nolimits}

\usepackage{algorithm}
\usepackage{algorithmic}

\begin{document}

\title{Statistical Anomaly Detection via Composite Hypothesis Testing
  for Markov Models$^*$ \thanks{* Submitted to IEEE Transactions on Signal Processing. Research partially supported by the
    NSF under grants CNS-1645681, CCF-1527292, and IIS-1237022, by the
    ARO under grants W911NF-11-1-0227 and W911NF-12-1-0390, and by a
    grant from the Boston Area Research Initiative (BARI).}}

\author{\IEEEauthorblockN{Jing Zhang$^\dag$, \IEEEmembership{Student
      Member, IEEE}, and Ioannis Ch. Paschalidis$^\ddag$,
    \IEEEmembership{Fellow, IEEE}} 
\thanks{\IEEEauthorblockA{$^\dag$ Division
of Systems Engineering, Boston University, Boston, MA 02446,
email: {\tt jzh@bu.edu}}.}
\thanks{$^\ddag$ Division of Systems Engineering, Dept. of Electrical
  and Computer Engineering, and Dept. of Biomedical Engineering, 
Boston University. 8 St. Mary's St., Boston, MA 02215, email: {\tt
 yannisp@bu.edu}, url: {\tt \url{http://sites.bu.edu/paschalidis}}.}
}

\maketitle
\begin{abstract}
  Under Markovian assumptions, we leverage a Central Limit Theorem (CLT)
  for the empirical measure in the test statistic of the composite
  hypothesis Hoeffding test so as to establish weak convergence results
  for the test statistic, and, thereby, derive a new estimator for the
  threshold needed by the test. We first show the advantages of our
  estimator over {an existing estimator} by conducting extensive
  numerical experiments. We find that our estimator controls better for
  false alarms while maintaining satisfactory detection
  probabilities. We then apply the Hoeffding test with our threshold
  estimator to detecting anomalies in two distinct applications domains:
  one in communication networks and the other in transportation
  networks. The former application seeks to enhance cyber security and
  the latter aims at building smarter transportation systems in cities.
\end{abstract}


\section{Introduction}

For a given system, {\em Statistical Anomaly Detection (SAD)} involves
learning from data the normal behavior of the system and
identifying/reporting time instances corresponding to atypical system
behavior. SAD has vast applications. For instance, motivated by the
importance of enhancing cyber security, recent literature has seen
applications in communication networks; see, e.g.,
\cite{pas-sma-ton-09,CDC09,wa-ro-ca-pa-anomaly-cdc13,robust-anomaly-tcns}. The
behavior of the system is typically represented as a time series of real
vectors and, in its most general version, anomaly detection is done
through some {\em Composite Hypothesis Test (CHT)}.

Specifically, a CHT aims to test the hypothesis that a given sequence of
observations is drawn from a known \textit{Probability Law (PL)} (i.e.,
\textit{probability distribution}) defined on a finite alphabet
\cite{TIT13}. Among numerous such tests, the one proposed by Hoeffding
\cite{hoeffding1965} has been well known for decades. When implementing
the Hoeffding test in the context of SAD, one must set appropriately a threshold $\eta$ so as to ensure a low false alarm rate while maintaining a reasonably high detection rate. In the
existing literature, this threshold is typically estimated by using Sanov's theorem
\cite{dembo1998large} -- a large deviations result. Note that such an
estimator (let us denote it by $\eta^{\text{sv}}$) is valid only in the
asymptotic sense. In practice, however, only a finite number of
observations are available, and it can be observed in simulations that
$\eta^{\text{sv}}$ is not accurate enough, especially for relatively small
sample sizes.

Our contributions in this paper include:

\begin{enumerate}
\item Under Markovian assumptions, we leverage a
  {\em Central Limit Theorem (CLT)} for a selected empirical measure
  related to the test statistic of the Hoeffding test, so as to establish weak convergence results for the test statistic, and derive a threshold estimator
  $\eta^{\text{wc}}$ therefrom, thus, extending the work of \cite{TIT13} which tackles
  the problem under independent and identically distributed (i.i.d.)
  assumptions.

\item We propose algorithms to calculate the threshold estimator
  $\eta^{\text{wc}}$ obtained above for the ordinary and a robust version of
  the Hoeffding test, respectively. We assess the advantages of our
  estimator over earlier work through numerical experiments. 

\item We apply the Hoeffding test with our threshold estimator to two
  types of systems for the purpose of anomaly detection: $(\rmnum{1})$ a
  communication network with flow data simulated by the software package
  SADIT \cite{SADIT}; and $(\rmnum{2})$ a real transportation network with jam
  data reported by Waze, a smartphone GPS navigation application. To the
  best of our knowledge, the latter is a novel application of anomaly
  detection.
\end{enumerate}

A preliminary conference version of this work appeared in
\cite{zh-pas-cdc15}. The present paper includes detailed technical
arguments, derives results for the robust version of the Hoeffding test,
expands the numerical comparisons with earlier work, and develops the
traffic jam anomaly detection application.

The rest of this paper is organized as follows. In Sec.~\ref{sec:liter}
we review related work. We formulate the threshold estimation problem in
Sec.~\ref{sec:prob} and derive theoretical results in
Sec.~\ref{sec:theory}. Sec.~\ref{sec:num} contains experimental
results. Concluding remarks are in Sec.~\ref{sec:con} and a number of
proofs appear in the Appendix.

\textbf{Notational conventions:} All vectors are column vectors. For
economy of space, we write $\bx = (x_1, \ldots, x_{\text{dim}(\bx)})$ to
denote the column vector $\bx$, where $\text{dim}(\bx)$ is its
dimension. We use prime to denote the transpose of a matrix or
vector. Denote by $\mathbb{N_+}$ the set of all nonnegative
integers. {$\|\bx\|$ denotes the $\ell_2$-norm of a vector $\bx$,
  $\lfloor x \rfloor$ the integer part of a positive number $x$, $|
  \scrA |$ the cardinality of a set $\scrA$}, $\log$ the natural
logarithm, $\mathbbm{P}(A)$ the probability of an event $A$,
$\mathbbm{E}[X]$ the expectation of a random variable $X$, and
$\text{Cov}(X_1, X_2)$ the covariance between two random variables $X_1$
and $X_2$. We use $\mathcal{N}(\bzero, \bSigma)$ to denote a Gaussian
distribution with zero mean and covariance matrix $\bSigma$. $X_1 \simeq
X_2$ indicates that the two random variables $X_1$ and $X_2$ have
approximately the same distribution. $\mathbbm{1}\{\cdot\}$ denotes the
indicator function and $\xrightarrow[{n \to \infty }]{\textit{d}}$
(resp., $\xrightarrow[{n \to \infty }]{{\textit{w.p.1}}}$) denotes
convergence \textit{in distribution} (resp., \textit{with probability
  one}) as $n$ approaches infinity.

\section{Related Work} \label{sec:liter} 

Modeling network traffic as stationary in time, \cite{pas-sma-ton-09}
applies two methods: one assumes the traffic to be an i.i.d.  sequence
and the other assumes observations of system activity follow a
finite-state Markov chain. Both methods are extended in
\cite{robust-anomaly-tcns} to the case where system activity is
time-varying. When implementing the Hoeffding test, however, both
\cite{pas-sma-ton-09} and \cite{robust-anomaly-tcns} use the large deviations estimator
$\eta^{\text{sv}}$ to calculate the detection threshold in a finite sample-size setting, thus controlling the false alarm rate not well enough.
 
To derive a more accurate threshold estimator,
\cite{TIT13,unnikrishnan2011universal} use a procedure commonly used by
statisticians: deriving results based on {\em Weak Convergence (WC)} of
the test statistic in order to approximate the error probabilities of
the Hoeffding test. Under i.i.d. assumptions, \cite{TIT13} (see also
\cite{unnikrishnan2011universal,wilks1938large}) proposes an alternative
estimator for $\eta$ (let us denote it by $\eta^{\text{wc}}$), which is
typically more accurate than $\eta^{\text{sv}}$, especially when not
that many samples are available.
 
There has also been work on obtaining a {tighter} approximation of
$\eta$ by refining Sanov's theorem \cite{iltis1995sharp}. However, such
refinements of large deviation results are typically faced with
computational difficulty; for instance, as noted in
\cite{unnikrishnan2011universal}, using the results of
\cite{iltis1995sharp} requires the computation of a surface integral.

Several alternative anomaly detection approaches have been proposed,
using for instance change detection
methods~\cite{thottan2003anomaly}. We refer the reader for a
comprehensive review of alternative methods to \cite{thottan2003anomaly}
and \cite{pas-sma-ton-09}.

\section{Problem Formulation} \label{sec:prob}

To model the statistical properties of a general system, we introduce a
few more notational conventions and some definitions. Let
$\boldsymbol{\Xi} = \{{\xi _i}; ~ i = 1, \ldots, N\}$ be a finite
alphabet containing $N$ symbols $\xi_1,\ldots,\xi_N$, and $\bY = \{
{Y_l};~l = 0,1,2, \ldots \}$ a time series of observations.  Define the
\textit{null hypothesis} $\cal H$ as: $\bY$ is drawn according to a
Markov chain with state set $\boldsymbol{\Xi}$ and transition matrix
$\bQ = \left[ {{q_{ij}}} \right]_{i,\,j = 1}^N$.  To further
characterize the {stochastic} properties of $\bY$, we define the
\textit{empirical Probability Law (PL)} by
\begin{equation} {\Gamma _n}( {{\theta_{ij}}} ) =
  \frac{1}{n}\sum_{l = 1}^n \mathbbm{1} \{ Z_l =
          \theta_{ij} \}, \label{1}
\end{equation}
where ${Z_l} = ( {{Y_{l - 1}},~{Y_l}}), ~l=1,\ldots,n$, ${\theta_{ij}} =
( {{\xi _i},~{\xi _j}}) \in \boldsymbol{\Xi} \times \boldsymbol{\Xi}$,
$i,\,j = 1, \ldots ,N$. Denote the transformed alphabet
${\bTheta} = \{ {{\theta_{ij}}};~i,\,j = 1, \ldots, N \} = \{
{{{\tilde {\theta}}_k}}; ~k = 1, \ldots, N^2 \}$ and note
${\bTheta} = \boldsymbol{\Xi} \times \boldsymbol{\Xi}$ with
${{\tilde {\theta}}_1} = {\theta_{11}}, \ldots, {{\tilde {\theta}}_N} =
{\theta_{1N}}, \ldots, {{\tilde {\theta}}_{( {N - 1})N + 1}} =
{\theta_{N1}}, \ldots, {{\tilde {\theta}}_{{N^2}}} = {\theta_{NN}}$.
Let also the set of PLs on ${\bTheta}$ be
$\mathcal{P}({\bTheta})$.

The transformed observations $\bZ = \{ {{Z_l};~l =
    1,2, \ldots } \}$ form a Markov chain evolving on
${\bTheta}$; denote its transition matrix by $\bP = [ {{p_{ij}}}
]_{i,\,j = 1}^{{N^2}}$ and the stationary distribution by
\begin{align}
{{\bpi}} = ( {{\pi _{ij}}};~i,\,j = 1, \ldots, N )
= ( {{{\tilde {\pi}}_k}};~k = 1, \ldots, N^2 ),   \label{pi-expre}
\end{align} 
where
${{\pi _{ij}}}$ denotes the probability of seeing $\theta_{ij}$, and
${{\tilde {\pi}}_1} = {{\pi}_{11}}, \ldots, {{\tilde {\pi}}_N} =
{{\pi}_{1N}}, \ldots, {{\tilde {\pi}}_{\left( {N - 1} \right)N + 1}}
= {{\pi}_{N1}}, \ldots, {{\tilde {\pi}}_{{N^2}}} = {{\pi}_{NN}}$.
We have \cite{dembo1998large}
\begin{align}
p( {{\theta_{ij}}\left| {{\theta_{kl}}} \right.} ) =
\mathbbm{1}{\{ {i = l} \}}{q_{ij}}, \quad k,\,l,\,i,\,j = 1,
\ldots ,N, \label{QtoP}
\end{align}  
which enables us to obtain $\bP$ directly from $\bQ$; see Remark \ref{re2} for
an example.  We can now restate the \textit{null hypothesis} $\mathcal{H}$ as:
the Markov chain $\bZ = \{ {{Z_l};~l = 1,2, \ldots }\}$ is drawn from PL
$\bpi$.

To quantify the distance between the empirical PL $\boldsymbol{\Gamma} _n$ and
the actual PL ${\bpi}$, one considers the \textit{relative entropy} (or
\textit{divergence}) between ${\boldsymbol{\Gamma} _n}$ and ${{\bpi} }$: 
\begin{equation}
  D( {\left. \boldsymbol{\Gamma} _n \right\| \bpi }) = 
 \sum_{i = 1}^N \sum_{j = 1}^N \Gamma_n(\theta_{ij}) \log 
   \frac{\Gamma _n(\theta_{ij})/
            \big( \tsum_{t = 1}^N \Gamma _n(\theta_{it})\big)} 
{\pi _{ij}/ \big(\tsum_{t = 1}^N \pi _{it} \big)},
\label{2}
\end{equation}
and the \textit{empirical measure}:
\begin{equation}
{\bU_n} = \sqrt n ( {{\boldsymbol{\Gamma} _n} - {\bpi} } ),
\label{2.0}
\end{equation}
where $\bpi$ is defined in (\ref{pi-expre}) and $\boldsymbol{\Gamma}_n$ is
  the vector
\[ {\boldsymbol{\Gamma} _n} = ({{\Gamma _n}( {{\theta_{11}}} )},
\ldots, {{\Gamma _n}( {{\theta_{1N}}} )}, \ldots, {{\Gamma
    _n} ( {{\theta_{N1}}} )}, \ldots, {{\Gamma _n}(
  {{\theta_{NN}}} )}).\]

Let now ${{\cal{H}}_n}$ be the output of a test that decides to accept
or to reject the \textit{null hypothesis} $\cal H$ based on the first
$n$ observations in the sequence $\bZ$.  Under Markovian assumptions
(Assumption \ref{ass1} in Sec. \ref{sec:theory}), the Hoeffding test
\cite{dembo1998large} is given by
\begin{equation}
{{\cal{H}}_n} ~\text{rejects}~ {\cal{H}} ~\text{if and only if}~ D( {\left. {{\boldsymbol{\Gamma} _n}} \right\|{{\bpi}}}) > \eta, \label{333}
\end{equation}
where $D( {\left. {{\boldsymbol{\Gamma} _n}} \right\|{{\bpi}}})$
(cf. \eqref{2}) is the \textit{test statistic} and $\eta$ is a
\textit{threshold}. 

{It is known that the Hoeffding test \eqref{333} satisfies
  asymptotic Newman-Pearson optimality
  \cite{pas-sma-ton-09,robust-anomaly-tcns}, in the sense that it
  maximizes the exponential decay rate of the \textit{misdetection
    probability} over all tests with a \textit{false positive
    probability} with exponential decay rate larger than $\eta$. Thus,
  an appropriate threshold $\eta$ should enable the test to have a small
  false positive rate while maintaining a satisfactorily high detection
  rate.}

{The theoretical \textit{false positive rate} \cite{TIT13}
of the test \eqref{333} is given by
\begin{equation}
{\beta} = {\mathbbm{P}_{\cal H}}( {D( {\left. {{\boldsymbol{\Gamma}
          _n}} \right\|{{\bpi}}} ) > \eta }),  \label{334}
\end{equation}
where the subscript $\cal H$ indicates that the probability is taken
under the null hypothesis.}

{Given a tolerable (target) $\beta$, by conducting an ROC analysis for the Hoeffding test using labeled training
  data, we could ``tune'' $\eta$ such that the corresponding \emph{discrete test}\footnote{A \emph{discrete test} corresponds to a fixed specific value for $\eta$ in \eqref{333}.} \cite{fawcett2006introduction} has a small false alarm rate and a high detection rate. In particular, we could select an $\eta$ corresponding to a point close to the northwest corner of the ROC graph. However, such tuning is too
  expensive and depends heavily on the quality and quantity of the
  training data. We can also, in principle, obtain the corresponding
  $\eta$ in \eqref{334} by directly simulating the samples of the test
  statistic $D( {\left. {{\boldsymbol{\Gamma}_n}} \right\|{{\bpi}}} ) $,
  thus deriving an empirical Cumulative Distribution Function (CDF) and
  using its $(1 - \beta)$-quantile. However, we will note in Remark
  \ref{re6} that this is also computationally too expensive when applied
  through a so-called ``windowing'' technique for purposes of anomaly
  detection. Thus, we seek to estimate $\eta$ without directly
  simulating the statistic. To that end, existing work involves using
  Sanov's theorem \cite{dembo1998large} to derive an estimator for
  $\eta$. Specifically, for large enough $n$, by replacing the right
  hand side in (\ref{334}) by an exponential we can obtain a minimal
  $\eta$ that suffices to bring the false positive rate below $\beta$
  \cite{pas-sma-ton-09,robust-anomaly-tcns}. Such an $\eta$ is given by}
\begin{equation}
  {\eta_{n,\beta}^{\text{sv}} \approx - (1/n) \log ( {{\beta}})}, \label{335}
\end{equation}
{where we use the $n,\beta$ subscript to denote the dependence of
  this estimator on $\beta$ and $n$ and the label $\text{sv}$ indicates
  that it is obtained from Sanov's theorem. We note that the estimator
  \eqref{335} does not contain any direct distributional information of
  the statistic $D( {\left. {{\boldsymbol{\Gamma} _n}}
    \right\|{{\bpi}}})$; this might be one of the causes leading to
  inaccurate estimation of $\eta_{n,\beta}$, especially when the sample
  size $n$ is relatively small in practice. To see this more clearly,
  one can consider an extreme scenario where $N = 4$, $\beta =
  10^{-1000}$, and $n = 50$ (this is a reasonably small value;
  comparable to $N^2 = 16$). Then by \eqref{335},
  $\eta_{n,\beta}^{\text{sv}}$ would be way bigger than necessary,
  tending to yield a test with zero false alarm rate but also zero
  detection rate for a typical test set. The issue arises because we use
  an asymptotic large deviations result for a relatively modest value of
  $n$.  Our primary goal in this paper is to derive an alternative
  threshold estimator, which would hopefully be more accurate than
  $\eta_{n,\beta}^{\text{sv}}$ for modest values of $n$, in terms of a
  certain metric that we will introduce in Sec.~\ref{sec:num}.}


\section{Theoretical Results} \label{sec:theory} 

We introduce the following assumption.
\begin{ass} \label{ass1} $\bZ= \{ {{Z_l}; ~l = 1,2, \ldots }\}$ is an
  aperiodic, irreducible, and positive recurrent Markov chain
  (\cite{jones2004markov}) evolving on ${\bTheta}$ with
  transition matrix $\bP$, stationary distribution ${\bpi}$, and with
  the same ${\bpi}$ as its initial distribution.
\end{ass}

\begin{rem} \label{re1} \emph{Since ${\bTheta}$ is a finite
    set, under Assumption~\ref{ass1} $\bZ$ is uniformly ergodic
    \cite{jones2004markov}. Assuming ${\bpi}$ as the initial
    distribution is done for notational simplicity; our results apply
    for any feasible initial distribution. Note also that, under
    Assumption \ref{ass1}, ${\bpi}$ must have full support over
    ${\bTheta}$; i.e., each entry in ${\bpi}$ is strictly
    positive.}
\end{rem}

\begin{lemma} \label{le1} Suppose Assumption \ref{ass1} holds. Then
\begin{equation}
  \frac{{{\pi _{ij}}}}{{\tsum_{t = 1}^N {{\pi _{it}}} }} = \frac{{{\pi _{ij}}}}{{\tsum_{t = 1}^N {{\pi _{ti}}} }} = {q_{ij}}, \quad i,\,j=1,\ldots,N.
\label{3}
\end{equation}
\end{lemma}
\begin{IEEEproof}
See Appendix \ref{sec:lem1}.
\end{IEEEproof}

\begin{rem} \label{re2} \emph{Under Assumption \ref{ass1}, Remark \ref{re1}
    and Lemma \ref{le1} imply that all entries of $\bQ$ are strictly
    positive, indicating that any two states of the original chain $\bY$
    are connected. This is a stringent condition; yet, in practice, if
    some $\pi_{ij}$ in \eqref{3} is zero, we can replace it with a small
    $\varepsilon > 0$, and then normalize the modified vector ${\bpi}$,
    thus ensuring that Assumption \ref{ass1} is satisfied.}

  \emph{Another reason why we set the zero entries in ${\bpi}$ to
    $\varepsilon > 0$ is for convenience of computing the original
    transition matrix $\bQ$, hence $\bP$, via \eqref{3} and
    \eqref{QtoP}. If we simply eliminate the corresponding states in
    $\bZ$, then it is possible that the number of the remaining
    states is not the square of some integer $N$; this would prevent us
    from easily recovering $\bP$ from ${\bpi}$. Consider the following
    example: Assuming
\[\bQ = \left[ {\begin{array}{*{20}{c}}
  {0.1}&{0.2}&{0.7} \\ 
  0&{0.2}&{0.8} \\ 
  {0.6}&{0.15}&{0.25} 
\end{array}} \right],\]
then by \eqref{QtoP} we have
\[\bP = \left[ {\begin{array}{*{20}{c}}
  {0.1}&{0.2}&{0.7}&0&0&0&0&0&0 \\ 
  0&0&0&0&{0.2}&{0.8}&0&0&0 \\ 
  0&0&0&0&0&0&{0.6}&{0.15}&{0.25} \\ 
  {0.1}&{0.2}&{0.7}&0&0&0&0&0&0 \\ 
  0&0&0&0&{0.2}&{0.8}&0&0&0 \\ 
  0&0&0&0&0&0&{0.6}&{0.15}&{0.25} \\ 
  {0.1}&{0.2}&{0.7}&0&0&0&0&0&0 \\ 
  0&0&0&0&{0.2}&{0.8}&0&0&0 \\ 
  0&0&0&0&0&0&{0.6}&{0.15}&{0.25} 
\end{array}} \right],
\]
and, by direct calculation, we obtain $\bpi = (0.03, 0.07, 0.23, 0,
\linebreak[3] 0.05, 0.14, 0.3, 0.07, 0.11)$.  Note that only $8$ entries
in $\bpi$ are non-zero and $8$ is not the square of some integer
$N$. Thus, if we eliminate the state corresponding to the zero entry in
$\bpi$, it will be hard to recover $\bQ$, hence $\bP$.}
\end{rem}

\subsection{Weak Convergence of Empirical Measure}
\label{WeakConvergenceOfUn}

Let us first establish CLT results for one-dimensional empirical measures
\begin{equation}
  U_{n,k} = \sqrt n ( {{\Gamma
        _n}( {{{\tilde {\theta}}_k}} ) - {{\tilde \pi }_k}}), 
\quad k = 1, \ldots ,{N^2}.
\label{2.1}
\end{equation}
For $k \in \{1,\ldots,N^2\}$ define
\begin{equation}
{f_k}( Z ) = \mathbbm{1}\{ {Z = {{\tilde {\theta}}_k}}
  \}.  
\label{998}
\end{equation}

\begin{lemma} \label{le2} Suppose Assumption \ref{ass1} holds. Then a
  Central Limit Theorem (CLT) holds for $U_{n,k}$; that is,
  $U_{n,k}\xrightarrow[{n \to \infty }]{{\textit{d}}}\mathcal{N}(
    {0,~\sigma _k^2})$ with $\sigma _k^2 = \operatorname{Cov}
  ( {{f_k} ( {{Z_1}}),~{f_k}( {{Z_1}})}
  ) + 2\sum_{m = 1}^\infty {\operatorname{Cov} } (
    {{f_k}( {{Z_1}} ),~{f_k}( {{Z_{1 + m}}} )}) < \infty$.
\end{lemma}
\begin{IEEEproof}
See Appendix \ref{sec:lem2}.
\end{IEEEproof}

{Now we state the CLT \cite[Thm. 3.1]{billingsley1961statisticala} for the multidimensional
empirical measure $\bU_n = (U_{n,k};\,k=1,\ldots,N^2)$ as Lemma \ref{th1}. Several different proofs for this result are available in \cite{billingsley1961statisticala} and the references therein. For completeness, we provide a proof that leverages the results from \cite{jones2004markov}, in terms of extending Lemma \ref{le2}.}

\begin{lemma} \label{th1} (\cite{billingsley1961statisticala}) Suppose Assumption \ref{ass1} holds. Then a
  multidimensional CLT holds for $\bU_n$; that is,
\begin{equation}
{\bU_n}\xrightarrow[{n \to \infty }]{{\textit{d}}}\mathcal{N}( {\bzero,
    \bLambda} ), 
\label{4}
\end{equation}
with $\bLambda = [{{\Lambda _{
          {ij} }}}]_{i,\,j = 1}^{{N^2}}$ being an $N^2 \times N^2$
    covariance matrix given by 
\begin{equation} {\Lambda _{ij}} = {{\tilde \pi }_i}( {{\bI_{ij}} -
        {{\tilde \pi }_j}} ) + \sum\limits_{m = 1}^\infty {[{{\tilde \pi
          }_i}( {\bP_{ij}^m - {{\tilde \pi }_j}} ) + {{\tilde \pi }_j}(
        \bP_{ji}^m - \tilde{\pi}_i )]}, \label{8}
\end{equation} 
where ${\bI}_{ij}$ denotes the $(i,\,j)$-th entry of the identity
matrix, and ${\bP}^m_{ij}$ (resp., ${\bP}^m_{ji}$) is the $(i,\,j)$-th
(resp., $(j,\,i)$-th) entry of the matrix ${\bP}^m$ (the $m$-th power of
${\bP}$), $i,\,j=1,\ldots,N^2$.
\end{lemma}
\begin{IEEEproof}
See Appendix \ref{sec:thm1}.
\end{IEEEproof}

\subsection{Weak Convergence of Test Statistic}

In this section, to derive weak convergence results for the test
statistic ${D}( {{\bnu} \|{\bpi} } )$, we will leverage a method
commonly-used by statisticians in terms of combining a Taylor's series
expansion for the test statistic and the CLT result for the empirical
measure \cite{wilks1938large}. Recently, under i.i.d. assumptions, such
a weak convergence analysis for certain test statistics has been
conducted in \cite{unnikrishnan2011universal,TIT13}.

To this end, for ${\bnu} \in
  \mathcal{P}( {\bTheta} )$ we consider
\begin{align}
  h( {\bnu} ) = {D}( {{\bnu} \|
        {\bpi} } ) = \sum\limits_{i = 1}^N
  {\sum\limits_{j = 1}^N {{\nu _{ij}}\log \frac{{\frac{{{\nu
                _{ij}}}}{{\sum\nolimits_{t = 1}^N {{\nu _{it}}}
            }}}}{{\frac{{{\pi _{ij}}}}{{\sum\nolimits_{t = 1}^N {{\pi
                  _{it}}} }}}}} }.
\label{90}
\end{align}
Let $\bU \sim \mathcal{N}( {\bzero, \bLambda } )$ with $\bLambda$ given
by \eqref{8}.  Now, we are in a position to derive weak convergence
results for our test statistic ${D}( {{\bnu} \|{\bpi} } )$.
\begin{thm} \label{th2} Suppose Assumption \ref{ass1} holds. Then we
  have the following weak convergence results: 
\begin{align}
D( \boldsymbol{\Gamma}_n \| \bpi ) &\xrightarrow[{n \to \infty
}]{\textit{d}}  \frac{1}{{2n}}{\bU^{\prime}}{\nabla ^2}h( {\bpi}
)\bU,  \label{336} \\ 
 D( \boldsymbol{\Gamma}_n \|
\boldsymbol{\pi} ) &\xrightarrow[{n \to \infty
}]{\textit{d}} \frac{1}{{2n}}\sum\limits_{k = 1}^{{N^2}} {{\rho _k}\chi
	_{1k}^2}, \label{chi2}
\end{align} 
where ${\nabla ^2}h( {\bpi} )$ is the Hessian of $h(\bnu)$ evaluated at
$\bnu = \bpi$, $\rho_k, \, k = 1, \ldots, N^2$, are the eigenvalues of
the matrix ${\nabla ^2}h( {\boldsymbol{\pi}} )\boldsymbol{\Lambda}$, and
$\chi _{1k}^2, \, k = 1, \ldots, N^2$, are $N^2$ independent $\chi^2$
random variables with one degree of freedom.
\end{thm}

\begin{IEEEproof}
	Let us first compute the gradient of $h(\bnu)$. Expanding the logarithm and after some algebra which leads to
cancellations of gradient terms with respect to $\nu _{ij}$ in $\sum_{t
  = 1}^N\nu _{it}$, for all $i,j = 1, \ldots ,N$, we obtain
\begin{equation}
\frac{{\partial h( {\boldsymbol{\nu }} )}}{{\partial {\nu _{ij}}}} =
\log {\nu _{ij}} - \log \bigg( {\sum\limits_{t = 1}^N {{\nu _{it}}} } \bigg) -
\log {\pi _{ij}}  
+ \log \bigg( {\sum_{t = 1}^N {{\pi _{it}}} } \bigg),  \label{92_updated}
\end{equation}
which implies
\begin{equation}
  \nabla h( {\bpi} ) = 0.
\label{93}
\end{equation}

Further, from \eqref{92_updated}, we compute the Hessian ${\nabla ^2}h(
{\bnu} )$ by
\begin{align}\frac{{{\partial ^2}h( {\bnu} )}}{{\partial {\nu
        _{ij}}\partial {\nu _{kl}}}} = \left\{ \begin{gathered} 
0,{\text{ if }}k \ne i, \hfill \\
\frac{1}{{{\nu _{ij}}}} - \frac{1}{{\sum_{t = 1}^N {{\nu _{it}}}
  }},{\text{ if }}k = i{\text{ and }}l = j, \hfill \\ 
- \frac{1}{{\sum_{t = 1}^N {{\nu _{it}}} }},{\text{ if }}k = i{\text{
    and }}l \ne j. \hfill \\  
\end{gathered}  \right. \label{hess}
\end{align}
Evaluating all the terms in \eqref{hess} at $\bnu = {\bpi}$ yields
${\nabla ^2}h( {\bpi})$, which will play a crucial role in approximating
${D}( {{\boldsymbol{\Gamma} _n} \| {{{\bpi}}} })$.
It is seen that ${\nabla
  ^2}h( {\bnu})$ is continuous in a neighborhood
of ${\bpi}$, and we can utilize the second-order Taylor's
series expansion of $h( {\bnu} )$ centered at
${\bpi}$ to express ${D}( {{\boldsymbol{\Gamma}
      _n}\| {{{\bpi}}}}) =
h(\boldsymbol{\Gamma} _n) - h( {\bpi}  )$. Specifically, by \eqref{93} and \eqref{2.0} we have
\begin{align}
2nD\left( {{\boldsymbol{\Gamma} _n}\left\| \bpi  \right.} \right) &= 2n\left( {h\left( {{\boldsymbol{\Gamma} _n}} \right) - h\left( \bpi  \right)} \right) \notag \\
&= n\left( {{\boldsymbol{\Gamma} _n} - \bpi } \right)'{\nabla ^2}h( {{{\tilde {\boldsymbol{\Gamma}} }_n}} )\left( {{\boldsymbol{\Gamma} _n} - \bpi } \right)  \notag \\
&= \bU_n'{\nabla ^2}h( {{{\tilde {\boldsymbol{\Gamma}} }_n}} )\bU_n, \label{94}
\end{align}
where ${{\tilde {\boldsymbol{\Gamma}} }_n} = {\xi _n}{\bGamma _n} + \left( {1 - {\xi _n}} \right)\bpi $ is determined with some ${\xi _n} \in [0,1]$. From the ergodicity of the chain $\bZ$ it follows ${\boldsymbol{\Gamma}
	_n}\xrightarrow[{n \to \infty }]{{\textit{w.p.1}}}{\bpi} $, leading to ${\tilde{\boldsymbol{\Gamma}}
	_n}\xrightarrow[{n \to \infty }]{{\textit{w.p.1}}}{\bpi} $. By the continuity of ${\nabla
	^2}h( {\bnu})$ we obtain 
\begin{align}
{\nabla ^2}h( {{{\tilde {\boldsymbol{\Gamma}} }_n}} ) \xrightarrow[{n \to \infty }]{{\textit{w.p.1}}} {\nabla ^2}h( {{{\bpi }}} ).  \label{slutsky}
\end{align}
 Applying Slutsky's theorem \cite{billingsley2013convergence}, by \eqref{4}, \eqref{94}, and \eqref{slutsky} we attain
\begin{equation}
D( \boldsymbol{\Gamma}_n \| \bpi ) =
\frac{1}{{2n}}{\bU_n^{\prime}}{\nabla ^2}h({{{\tilde {\boldsymbol{\Gamma}} }_n}} )\bU_n \xrightarrow[{n \to \infty
}]{\textit{d}}  \frac{1}{{2n}}{\bU^{\prime}}{\nabla ^2}h( {\bpi} )\bU.  \notag
\end{equation}

Finally, by means of a linear transformation
	\cite{imhof1961computing} on the quadratic form ${\bU^{\prime}}{\nabla ^2}h( {\bpi} )\bU$, we derive the following alternative
	asymptotic result: 
	\begin{align} D( \boldsymbol{\Gamma}_n \|
	\boldsymbol{\pi} ) = \frac{1}{{2n}}{\bU_n^{\prime}}{\nabla ^2}h({{{\tilde {\boldsymbol{\Gamma}} }_n}} )\bU_n \xrightarrow[{n \to \infty
	}]{\textit{d}} \frac{1}{{2n}}\sum\limits_{k = 1}^{{N^2}} {{\rho _k}\chi
		_{1k}^2}, \notag
	\end{align}
	where $\rho_k, \, k = 1, \ldots, N^2$, are the eigenvalues of the matrix
	${\nabla ^2}h( {\boldsymbol{\pi}} )\boldsymbol{\Lambda}$, and $\chi
	_{1k}^2, \, k = 1, \ldots, N^2$, are $N^2$ independent $\chi^2$ random
	variables with one degree of freedom.
\end{IEEEproof}

\subsection{Threshold Approximation}

We use an empirical Cumulative Distribution Function (CDF) to
approximate the actual CDF of $D( \boldsymbol{\Gamma}_n \|\bpi)$. In particular, it is seen from \eqref{336} that ${D}(
{{\boldsymbol{\Gamma} _n}\| {\bpi} } ) \simeq
(1/(2n)) {\bU^{\prime}}{\nabla ^2}h( {\bpi} )\bU$ for large $n$.
Thus, to derive an empirical CDF of ${D}( {{\boldsymbol{\Gamma} _n}\|
  {\bpi} } )$, we can generate a set of Gaussian sample vectors
independently according to $\mathcal{N}( {\bzero, \bLambda } )$ and then
plug each such sample vector into the right-hand side of \eqref{336}
(i.e., replace $\bU$), thus, obtaining a set of sample scalars, as a
reliable proxy for samples of ${D}( {{\boldsymbol{\Gamma} _n}\| {\bpi} }
)$.

Once we obtain an empirical CDF of ${D}( {{\boldsymbol{\Gamma}
      _n}\| {\bpi} } )$, say, denoted
${F_{\text{em}}}( \cdot ; n )$, then, by \eqref{334}, we can
estimate $\eta_{n,\beta}$ as
\begin{align}
\eta_{n,\beta}^{\text{wc}}  \approx F_{\text{em}}^{ - 1}( {1 - {\beta}}; n ),
\label{337}
\end{align}
where $F_{\text{em}}^{-1}(\cdot; n)$ is the inverse of
$F_{\text{em}}(\cdot; n)$. {Note that, the
  $\eta_{n,\beta}^{\text{wc}}$ derived by \eqref{337} depends on the
  entries of the PL $\bpi$. In practice, if ${\bpi}$ is not directly
  available, we can replace it by the empirical PL evaluated over a long
  past sample path. For such cases, we summarize the procedures of
  estimating the threshold based on our weak convergence analysis as
  Alg.~\ref{alg:thres}, where $\hat \bpi$ is a good estimate for
  ${\bpi}$. We note that the length $n_0$ of the past sample path should
  be sufficiently large (e.g., $n_0 \ge 500N^2$) so as to guarantee the
  validity of taking $\bpi$ to be $\hat \bpi$. In addition, the small
  positive number $\varepsilon$ (e.g., $\varepsilon \le 10^{-6}$)
  introduced in Step 1 is to avoid division by zero, thus ensuring the
  numerical stability of the algorithm. If, on the other hand, the
  actual PL $\bpi$ is known, then we can still apply
  Alg. \ref{alg:thres} by replacing the $\hat \bpi$ therein with
  $\bpi$.}

 {Similar to \eqref{337}, we
can derive another weak convergence based threshold estimator
$\bar{\eta}_{n,\beta}^{\text{wc}}$ from \eqref{chi2}. However, an easy
way of calculating $\bar{\eta}_{n,\beta}^{\text{wc}}$ (also summarized
in Alg. \ref{alg:thres}) still cannot avoid simulations; it is hard to
conclude any advantage of $\bar{\eta}_{n,\beta}^{\text{wc}}$ over
${\eta}_{n,\beta}^{\text{wc}}$. As a matter of fact, calculating the
eigenvalues of ${\nabla ^2}h( {\boldsymbol{\pi}} )\boldsymbol{\Lambda}$
makes the calculation of $\bar{\eta}_{n,\beta}^{\text{wc}}$ numerically
not as stable, compared to the calculation of
${\eta}_{n,\beta}^{\text{wc}}$ via Alg. \ref{alg:thres}. Other methods
for numerically obtaining $\bar{\eta}_{n,\beta}^{\text{wc}}$ can be
found, e.g., in \cite{liu2009new} and the references therein. Another
fact we should point out is that, in
\cite[p. 30]{billingsley1961statistical}, a slightly different statistic
is considered and therefore an even simpler asymptotic distribution can
be derived correspondingly. Moreover, some other papers, e.g.,
\cite{menendez1997testing,menendez1999inference}, also considered
similar but different statistics.}

We will illustrate by extensive experiments that our weak convergence
analysis can empirically produce more accurate estimation of the
threshold than Sanov's theorem {for moderate values of $n$}; the
price we have to pay, however, is a relatively long but still acceptable
computation time.

\begin{algorithm}
	\caption{Threshold estimation for the ordinary Hoeffding test
          under Markovian assumptions based on weak convergence analysis.}
	\label{alg:thres}
	\begin{algorithmic}[1]
          \REQUIRE {The sample size} $n$, the target false
          positive rate $\beta$, the alphabet ${\bTheta} = \{
          {{{\tilde {\theta}}_k}}; ~k = 1, \ldots, N^2 \}$,
          a sample path of the chain $\bZ$, denoted
          ${\bZ^{\scriptscriptstyle{( 0 )}}} = \{
          {Z_1^{\scriptscriptstyle{( 0 )}}, \ldots
            ,Z_{n_0}^{\scriptscriptstyle{( 0 )}}} \}$, where $n_0$
          is the length, {and the Boolean parameter $\chi^2_\text{enab}$}.
\STATE Estimate $ {{{\tilde {\pi}}_k}}$ by
\[ 
{{\hat {\tilde {\pi}} }_k} = \max \bigg\{
		{\frac{1}{n_0}{\sum\limits_{i = 1}^{{n_0}} {\mathbbm{1}{\big\{
	{Z_i^{\scriptscriptstyle{( 0 )}} = {{{\tilde {\theta}}_k}}} \big\}}}
			}, ~\varepsilon } \bigg\},\; k=1,\ldots,N^2,
\]
where $\varepsilon > 0$ is a small number.  
\STATE Estimate ${\bpi}$ as
$\hat {{\bpi}} = ( {{{{{\hat{ \tilde {\pi}} }_k}}} \mathord{/
    {\vphantom {1 2}} \kern-\nulldelimiterspace} {{\hat s}}}; ~k = 1,
\ldots, N^2 )$, where $\hat s = \sum_{j = 1}^{{N^2}}
{{{\hat {\tilde{ \pi }}}_j}} $ is a normalizing constant.  
\STATE Estimate ${\nabla ^2}h( {\bpi} )$ as ${\nabla
  ^2}h( \hat {{\bpi}} )$, by plugging $\hat {{\bpi}}$ into \eqref{hess}
(i.e., using $\hat {{\bpi}}$ to replace ${\bnu}$).  
\STATE Estimate
$\bP$ as $\hat {\bP}$, via (cf. \eqref{QtoP} and Lemma \ref{le1})
\[
{\hat p}( {{\theta_{ij}}| {{\theta_{kl}}} } ) =
\mathbbm{1}{\{ {i = l} \}}{{\hat q}_{ij}}, \quad k,l,i,\,j = 1, \ldots ,N, 
\]
where 
$ {\hat q}_{ij} = {\hat \pi}_{ij}/(\sum_{t = 1}^N {\hat \pi }_{it})$.
\STATE Estimate $\bLambda$ as $\hat {\bLambda}$, using (by \eqref{8} in
Lemma \ref{th1}) 
\[
{{\hat \Lambda }_{ij}} = {{\hat {\tilde \pi} }_i} ( {{{\mathbf{I}}_{ij}} - {{\hat {\tilde \pi }}_j}} ) + \sum\limits_{m = 1}^{{m_0}} \Big[{{\hat {\tilde \pi }}_i}( {\hat {\mathbf{P}}_{ij}^m - {{\hat {\tilde \pi} }_j}} ) 
+ {{\hat {\tilde \pi} }_j}( {\hat {\mathbf{P}}_{ji}^m - {{\hat {\tilde
        \pi} }_i}} )\Big], 
\]
where $m_0$ is a sufficiently large integer.	
\STATE Update  $\hat {\bLambda}$ by setting ${(\hat{\bLambda} +
\hat{\bLambda}^{\prime}) \mathord{/ {\vphantom {1 2}} 
\kern-\nulldelimiterspace} 2}$ to $\hat {\bLambda}$.
{\IF{$\chi^2_\text{enab} = \text{FALSE}$}
\STATE Generate $T$ Gaussian sample vectors ${\hat
  {\bU}}^{\scriptscriptstyle{(t)}}, \,t=1,\ldots,T$, according to
$\mathcal{N}( {\bzero, \hat{\bLambda} } )$. 
\STATE Estimate $T$ samples of ${D}({{\boldsymbol{\Gamma} _n}\| {\bpi} })$ as 
$(1/(2n)) {\hat {\bU}^{\scriptscriptstyle{( t )}'}} {\nabla ^2}h( {\hat
  {{\bpi}} }) {\hat {\bU}^{\scriptscriptstyle{( t )}}}$, $t = 1, \ldots
,T$ (cf. \eqref{336}). 
\STATE Based on the $T$ samples obtained in the last step, estimate an
empirical CDF of ${D}({{\boldsymbol{\Gamma} _n}\| {{{\bpi}}} })$,
denoted ${F_{\text{em}}}( \cdot; n )$. 
\STATE Obtain an estimated value for $\eta_{n,\beta}$ by calculating
$\eta_{n,\beta}^{\text{wc}}$ via \eqref{337}. 
\ELSIF{$\chi^2_\text{enab} = \text{TRUE}$}
\STATE Calculate the eigenvalues $\hat \rho_k, \, k = 1, \ldots, N^2$, of the matrix ${\nabla ^2}h( \hat{\boldsymbol{\pi}} )\hat {\boldsymbol{\Lambda}}$. 
\STATE Generate $T$ samples of $(1/(2n))\sum\nolimits_{k = 1}^{{N^2}} {{\hat \rho _k}\chi _{1k}^2}$ (cf. \eqref{chi2}). 
\STATE Based on the $T$ samples obtained in the last step, estimate an
empirical CDF of ${D}({{\boldsymbol{\Gamma} _n}\| {{{\bpi}}} })$,
denoted  ${\bar F_{\text{em}}}( \cdot; n )$. 
\STATE Obtain an estimated value for $\eta_{n,\beta}$ by calculating
$\bar \eta_{n,\beta}^{\text{wc}}$ via \eqref{337} with
$F_{\text{em}}(\cdot; n)$ replaced by $\bar F_{\text{em}}(\cdot; n)$.
\ENDIF}
\end{algorithmic}
\end{algorithm}

\begin{rem} \label{rem:alg} \emph{In Alg. \ref{alg:thres}, due to
    acceptable numerical errors, the originally estimated $\hat{\bLambda}$
    (Step 5) could be neither symmetric nor positive
    semi-definite. Symmetry is imposed by Step 6. Further, to ensure
    positive semi-definiteness we can diagonalize $\hat
    \bLambda$ as
\begin{equation}
\hat {\bLambda}  = {{\bO}^{ - 1}}  \text{diag}(\lambda_1,\ldots, \lambda _{N^2}) 
\bO, \label{qrf}
\end{equation}
where $\bO$ is an orthogonal matrix and $\text{diag}(\blambda)$ a
diagonal matrix with the elements of $\blambda$ in the main diagonal.
Due to numerical errors we might encounter cases where some $\lambda_i$
are either negative or too small; we can replace them with small
positive numbers and recalculate the right-hand side of \eqref{qrf},
thus obtaining an updated positive-definite $\hat \bLambda$.  For
implementation details, the reader is referred to \cite{TAHTMA}.  }
\end{rem}

\subsection{A Robust Hoeffding Test}

Many actual systems exhibit time-varying behavior. In this section, we
extend our methodology to accommodate such systems and use a set of PLs
(instead of a single PL $\bpi$) to model past system activity.

Let the \textit{null hypothesis} $\mathcal{H}$ be defined as: $\bZ = \{
{{Z_l};~l = 1,2, \ldots } \}$ is drawn according to the set of PLs
$\boldsymbol{\Pi} = \{\bpi^{\scriptscriptstyle{(1)}}, \ldots,
\bpi^{\scriptscriptstyle{(L)}}\} \subset
\mathcal{P}({\bTheta})$, i.e., $\bZ$ is drawn from one of the
PLs in $\boldsymbol{\Pi}$ but we do not know from which one. Consider a
robust version of the Hoeffding test
\cite{robust-anomaly-tcns,TIT13,pandit2006worst} under Markovian
assumptions:
\begin{equation} {{\cal{H}}_n}{\text{ rejects }}{\cal{H}}{\text{ if and
      only if }}\mathop {\inf }_{\bpi \in \boldsymbol{\Pi} } D(
  {{\boldsymbol{\Gamma} _n}\| \bpi } ) > {\eta}. \label{robust-ht}
\end{equation}
Essentially, the test selects the most likely PL from $\boldsymbol{\Pi}$
and uses that to make a decision as in (\ref{333}). Asymptotic
Newman-Pearson optimality of this test is shown in
\cite{robust-anomaly-tcns}.

For $l = 1,\ldots, L$, let $\bP^{{\scriptscriptstyle(l)}}$ denote the
transition matrix corresponding to $\bpi^{\scriptscriptstyle{(l)}}$ and,
similar to \eqref{pi-expre}, we write
\[
{{\bpi}^{\scriptscriptstyle{(l)}}} = ( {{\pi^{\scriptscriptstyle{(l)}}
    _{ij}}}; i,j = 1, \ldots, N )
= ( {{{\tilde {\pi}}^{\scriptscriptstyle{(l)}}_k}}; k = 1, \ldots, N^2 ).   
\]
Assume $\bZ$ is drawn from PL $\bpi^{\scriptscriptstyle{(l)}}$ which
satisfies Assumption \ref{ass1}. Let $\bU_n^{\scriptscriptstyle{(l)}} =
\sqrt n ( {{\boldsymbol{\Gamma} _n} - {\bpi ^{\scriptscriptstyle{(l)}}}}
)$. By Lemma \ref{th1}, we have
\begin{align}
\bU_n^{\scriptscriptstyle{(l)}} \xrightarrow[{n \to \infty
}]{{\textit{d}}}{\cal{N}}( {\bzero, {\boldsymbol{\Lambda} ^{\scriptscriptstyle{(l)}}}} ),  \label{clt-robust}
\end{align} 
where $\bLambda^{\scriptscriptstyle{(l)}} = \big[\Lambda ^{\scriptscriptstyle{(l)}}_
{ij}\big]_{i,\,j = 1}^{{N^2}}$ is given by
\begin{align}
\Lambda ^{\scriptscriptstyle{(l)}}_{ij} ~=~& {{\tilde \pi }_i^{\scriptscriptstyle{(l)}}}( {{\bI_{ij}} - {{\tilde \pi }_j^{\scriptscriptstyle{(l)}}}} ) \nonumber \\
&+ \sum\limits_{m = 1}^\infty  {[{{\tilde \pi }_i^{\scriptscriptstyle{(l)}}}( {\bP_{ij}^{\scriptscriptstyle{(l)}m} - {{\tilde \pi }_j^{\scriptscriptstyle{(l)}}}} ) 
	+ {{\tilde \pi }_j^{\scriptscriptstyle{(l)}}}( \bP_{ji}^{\scriptscriptstyle{(l)}m} - \tilde{\pi}_i^{\scriptscriptstyle{(l)}} )]},  \notag 
\end{align} 
with $\bP_{ij}^{\scriptscriptstyle{(l)}m}$ being the $(i,j)$-th entry of
the matrix $\bP^{\scriptscriptstyle{(l)}m}$ (the $m$-th power of
$\bP^{\scriptscriptstyle{(l)}}$). Let $\bU^{\scriptscriptstyle{(l)}}
\sim {\cal{N}}( {\bzero, {\boldsymbol{\Lambda} ^{\scriptscriptstyle{(l)}}}} )
$. Using \eqref{336} we obtain
\[ {D}( {{\boldsymbol{\Gamma} _n}\| {\bpi}^{\scriptscriptstyle{(l)}} } )
\simeq \frac{1}{{2n}}{\bU^{\scriptscriptstyle{(l)}\prime}}{\nabla
  ^2}h( {\bpi}^{\scriptscriptstyle{(l)}}
)\bU^{\scriptscriptstyle{(l)}},
\]
which leads to an approximation for the infimum term in \eqref{robust-ht}:
\begin{align}
\mathop {\inf }\limits_{\bpi  \in \boldsymbol{\Pi} } D( {{\boldsymbol{\Gamma} _n}\| \bpi  } ) \simeq \mathop {\inf }\limits_{l  \in \{1, \ldots, L\} } \frac{1}{{2n}}{\bU^{\scriptscriptstyle{(l)}\prime}}{\nabla
  ^2}h( {\bpi}^{\scriptscriptstyle{(l)}} )\bU^{\scriptscriptstyle{(l)}}.
  \label{approx-11}
\end{align}
By the right-hand side of \eqref{approx-11}, we can generate Gaussian
samples to compute a reliable proxy for the samples of $\mathop {\inf
}_{\bpi \in \boldsymbol{\Pi} } D( {{\boldsymbol{\Gamma} _n}\| \bpi } )$,
thereby, obtaining an empirical CDF, denoted ${F^{\text{rob}}_{\text{em}}}( \cdot ; n
)$, of $\mathop {\inf }_{\bpi \in \boldsymbol{\Pi} } D(
{{\boldsymbol{\Gamma} _n}\| \bpi } )$. Thus, given a target false
positive rate {$\beta$}, similar to \eqref{337}, we can estimate the
threshold $\eta_{n,\beta}$ as
\begin{align}
\eta_{n,\beta}^{\text{wc}}  \approx (F_{\text{em}}^{\text{rob}})^{ - 1}( {1 - {\beta}}; n ),
\label{eta-wc-rob}
\end{align}
where $(F_{\text{em}}^{\text{rob}})^{ - 1}( \cdot ; n )$ denotes the inverse of
${F^{\text{rob}}_{\text{em}}}( \cdot ; n )$. {Similar to \eqref{chi2}, we can also derive a $\chi^2$-type asymptotic approximation to the distribution of $\mathop {\inf }\nolimits_{\bpi  \in \boldsymbol{\Pi} } D( {{\boldsymbol{\Gamma} _n}\| \bpi  } ) $, thus obtaining another WC-based threshold estimator $\bar \eta_{n,\beta}^{\text{wc}} $; for economy of space, we omit the details.} For the cases where the PLs are not directly available, we summarize the calculation of $\eta_{n,\beta}^{\text{wc}} $ for the robust Hoeffding test as
Alg. \ref{alg:thresrob1}.

\begin{algorithm}
  \caption{Threshold estimation for the robust Hoeffding test under
    Markovian assumptions based on weak convergence analysis.}
	\label{alg:thresrob1}
\begin{algorithmic}[1]
  \REQUIRE The sample size $n$, the target false positive
  rate $\beta$, the alphabet ${\bTheta} = \{
  {{{\tilde {\theta}}_k}}; ~k = 1, \ldots, N^2 \}$, and a sample path
  of each PL $\bpi^{\scriptscriptstyle{(l)}}$, denoted
  ${\bZ^{\scriptscriptstyle{( l0 )}}} = \{ {Z_1^{\scriptscriptstyle{( l0
        )}}, \ldots, Z_{n_0}^{\scriptscriptstyle{( l0 )}}}\}$, where
  $n_0$ is the length, $l=1,\ldots,L$.  

\FOR {$l=1,\ldots,L$}

\STATE \label{line2} Estimate $ {{{\tilde
      {\pi}}_k^{\scriptscriptstyle{(l)}}}}$, $k=1,\ldots,N^2$, by
\[
{{\hat {\tilde {\pi}} }_k^{\scriptscriptstyle{(l)}}} = \max \bigg\{
\frac{1}{n_0} \sum_{i = 1}^{n_0} \mathbbm{1}\{
Z_i^{\scriptscriptstyle{(l0)}} = \tilde{\theta}_k \}, \varepsilon \bigg\},
\]
where $\varepsilon > 0$ is a small number. 

\STATE \label{line3} Estimate ${\bpi}^{\scriptscriptstyle{(l)}}$ as
${\hat{{\bpi}}^{\scriptscriptstyle{(l)}}} = ( {{{{{\hat{ \tilde
            {\pi}}}_k^{\scriptscriptstyle{(l)}}}}} \mathord{/ {\vphantom
      {1 2}} \kern-\nulldelimiterspace} {{\hat
      s^{\scriptscriptstyle{(l)}}}}}; ~k = 1, \ldots, N^2 )$, where
$\hat s^{\scriptscriptstyle{(l)}} = \sum\nolimits_{j = 1}^{{N^2}}
{{{\hat {\tilde{ \pi }}}_j^{\scriptscriptstyle{(l)}}}} $ is normalizing
constant.

\STATE Estimate ${\nabla^2}h( {\bpi}^{\scriptscriptstyle{(l)}} )$ as
${\nabla^2}h( \hat {{\bpi}}^{\scriptscriptstyle{(l)}} )$, by plugging
$\hat {{\bpi}}^{\scriptscriptstyle{(l)}}$ into \eqref{hess} (i.e., using
$\hat {{\bpi}}^{\scriptscriptstyle{(l)}}$ to replace ${\bnu}$).  

\STATE Estimate $\bP^{\scriptscriptstyle{(l)}}$ as $\hat
{\bP}^{\scriptscriptstyle{(l)}}$, via (cf. \eqref{QtoP} and Lemma
\ref{le1})
\[ {\hat p^{\scriptscriptstyle{(l)}}}( {{\theta_{ij}}| {{\theta_{kl}}} } ) =
\mathbbm{1}{\{ {i = l} \}}{{\hat q}_{ij}^{\scriptscriptstyle{(l)}}},
\quad k,l,i,\,j = 1, \ldots ,N, 
\]
where $ {{\hat q}_{ij}^{\scriptscriptstyle{(l)}}} = {\hat
  \pi}^{\scriptscriptstyle{(l)}}_{ij}/(\sum_{t = 1}^N {\hat
  \pi}^{\scriptscriptstyle{(l)}}_{it}).$ 

\STATE Estimate $\bLambda^{\scriptscriptstyle{(l)}}$ as $\hat
{\bLambda}^{\scriptscriptstyle{(l)}}$, using (by \eqref{8} in
Lemma \ref{th1})
\begin{align*}
{{\hat \bLambda }^{\scriptscriptstyle{(l)}}_{ij}} = & {{\hat {\tilde \pi} }^{\scriptscriptstyle{(l)}}_i}( {{{\mathbf{I}}_{ij}} - {{\hat {\tilde \pi }}^{\scriptscriptstyle{(l)}}_j}} ) + \sum\limits_{m = 1}^{{m_0}} {} \Big[{{\hat {\tilde \pi }}^{\scriptscriptstyle{(l)}}_i}( {\hat {\mathbf{P}}_{ij}^{\scriptscriptstyle{(l)}m} - {{\hat {\tilde \pi} }^{\scriptscriptstyle{(l)}}_j}} ) \\
				&+ {{\hat {\tilde \pi} }^{\scriptscriptstyle{(l)}}_j}( {\hat {\mathbf{P}}_{ji}^{\scriptscriptstyle{(l)}m} - {{\hat {\tilde \pi} }^{\scriptscriptstyle{(l)}}_i}} )\Big], 
\end{align*}
where $m_0$ is a sufficiently large integer.

\STATE Update  $\hat {\bLambda}^{\scriptscriptstyle{(l)}}$ by setting
${(\hat{\bLambda}^{\scriptscriptstyle{(l)}} + \hat{\bLambda}^{\scriptscriptstyle{(l)}'}) \mathord{/ {\vphantom {1 2}} \kern-\nulldelimiterspace} 2}$ to $\hat {\bLambda}^{\scriptscriptstyle{(l)}}$.
				
\STATE Generate $T$ Gaussian sample vectors ${\hat
  {\bU}}^{\scriptscriptstyle{(lt)}}, \,t=1,\ldots,T$, according to
$\mathcal{N}( {\bzero, \hat{\bLambda}^{\scriptscriptstyle{(l)}} } )$.

\ENDFOR

\STATE Estimate $T$ samples of $\mathop {\inf }_{\bpi \in
  \boldsymbol{\Pi} } D( {{\boldsymbol{\Gamma} _n}\| \bpi } )$ as
$\mathop {\inf }_{l \in \{1, \ldots, L\} } (1/2n) {\hat
  \bU^{\scriptscriptstyle{(lt)}\prime}} {\nabla ^2}h(
{\bpi}^{\scriptscriptstyle{(l)}} )\hat \bU^{\scriptscriptstyle{(lt)}}$,
$t = 1, \ldots ,T$ (cf. \eqref{approx-11}).
				
\STATE Based on the $T$ samples obtained in the last step, estimate an
empirical CDF of $\mathop {\inf }_{\bpi \in \boldsymbol{\Pi} } D(
{{\boldsymbol{\Gamma} _n}\| \bpi } )$, denoted ${F^{\text{rob}}_{\text{em}}}( \cdot ;
n )$.

\STATE Obtain an estimated value for $\eta_{n,\beta}$ by calculating
$\eta_{n,\beta}^{\text{wc}}$ via \eqref{eta-wc-rob}.
\end{algorithmic}
\end{algorithm}

\begin{figure}[thpb]  
	\centering
	\begin{subfigure}[b]{0.485\textwidth}
		\includegraphics[width=\textwidth]{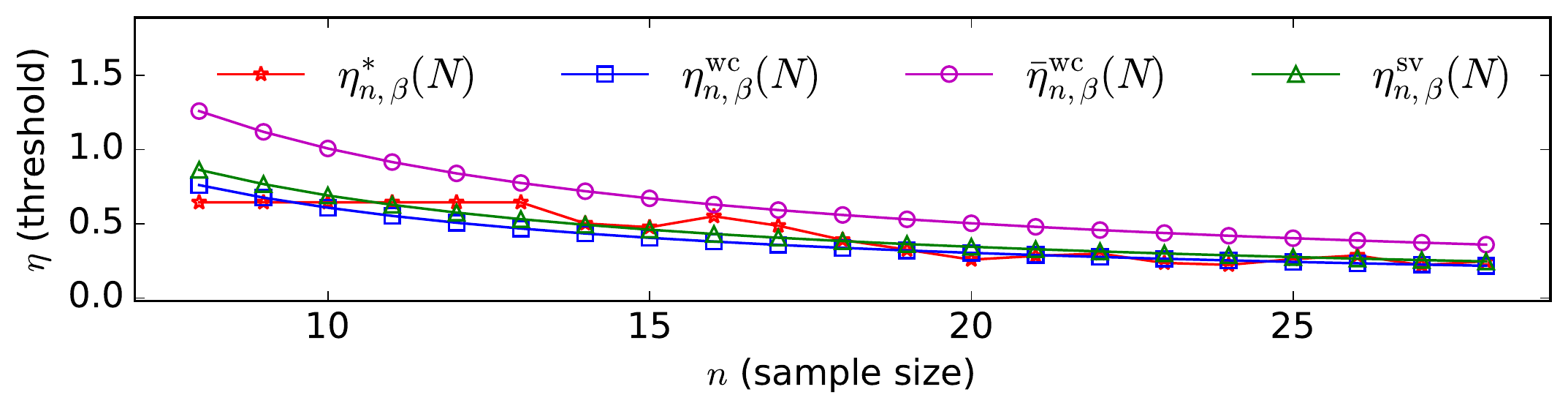} 
		\caption{$N = 2$.}
		\label{fig1:N2}
	\end{subfigure} 
	\begin{subfigure}[b]{0.485\textwidth}
		\includegraphics[width=\textwidth]{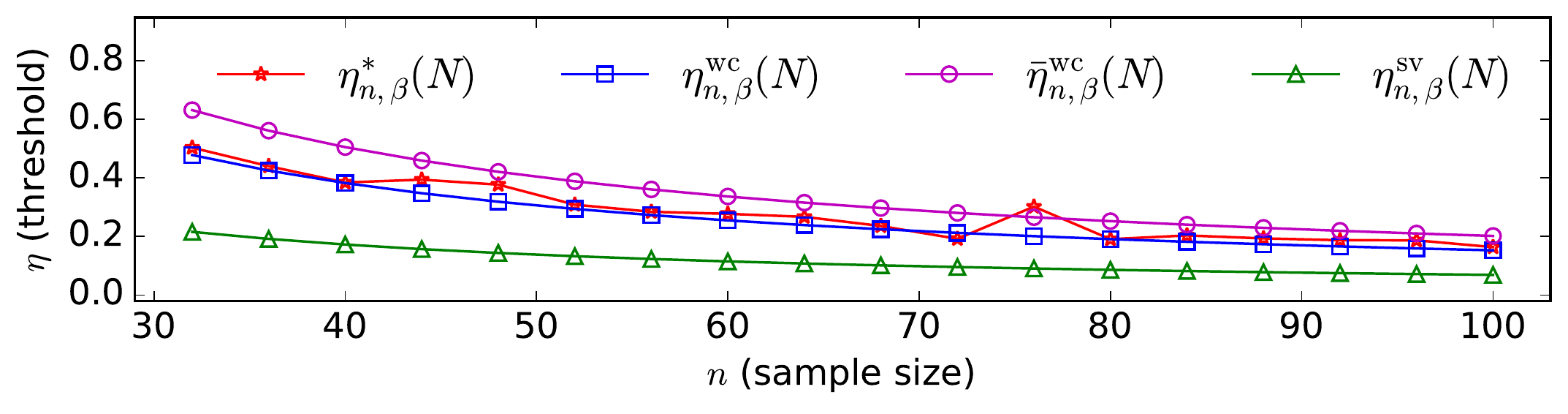}
		\caption{$N = 4$.}
		\label{fig1:N3}
	\end{subfigure}   
	\begin{subfigure}[b]{0.485\textwidth}
		\includegraphics[width=\textwidth]{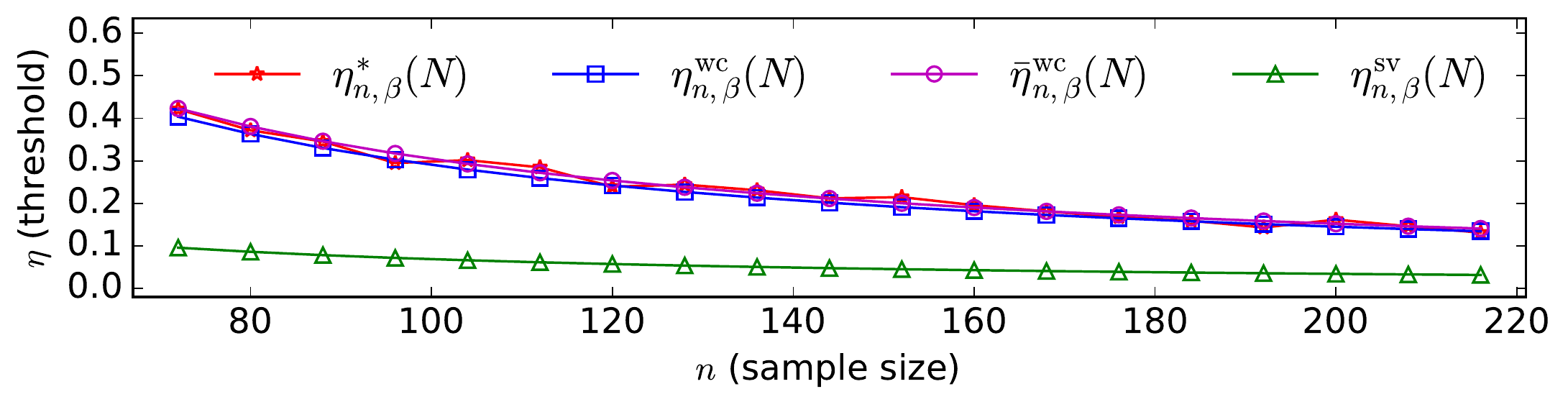}
		\caption{$N = 6$.}
		\label{fig1:N4}
	\end{subfigure}  
	\begin{subfigure}[b]{0.485\textwidth}
		\includegraphics[width=\textwidth]{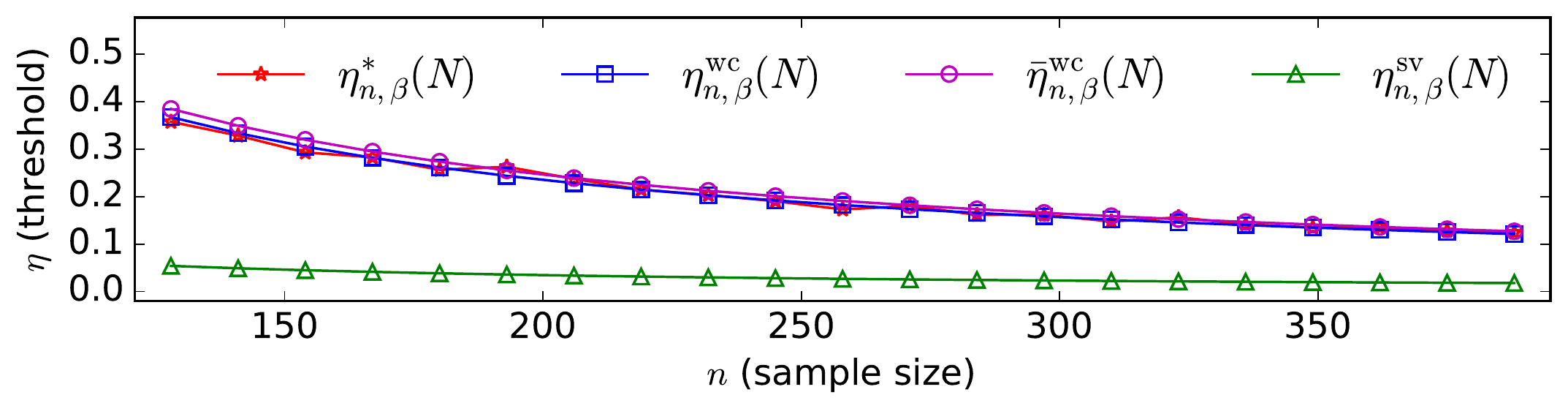}
		\caption{$N = 8$.}
		\label{fig1:N5}
	\end{subfigure}  
	\caption{Threshold versus sample size; scenarios
		corresponding to $\beta = 0.001$, $N = 2, \,4, \,6, \,8$.}
	\label{fig1:N2345}
\end{figure}

\section{Experimental Results} \label{sec:num}

In this section, we assess the accuracy of our threshold estimator and
the performance of the anomaly detection procedure. We start with a
numerical evaluation of the threshold's accuracy and then perform
anomaly detection in two application settings using simulated and actual
data.

\subsection{Numerical Results for Threshold
  Approximation} \label{sec:num.A} 

In this subsection, for simplicity we consider the ordinary (and not the
robust) Hoeffding test. We have developed a software package TAM
\cite{TAHTMA} to perform the experiments. We will use
${\bTheta} = \{1,2,\ldots,N^2\}$ to indicate the states and
assume the stationary distribution ${\bpi}$ to also be the initial
distribution.

In the following numerical examples, we first randomly create a valid
(i.e., such that Assumption \ref{ass1} holds) $N \times N$ transition
matrix $\bQ$, giving rise to an $N^2 \times N^2$ transition matrix
$\bP$, and then generate $T$ {test} sample paths of the chain $\bZ$,
each with length $n$, denoted ${\bZ^{\scriptscriptstyle{( t )}}} = \{
{Z_1^{\scriptscriptstyle{( t )}}, \ldots, Z_n^{\scriptscriptstyle{( t
      )}}}\}$, $t = 1, \ldots ,T$. We use these samples to derive
empirical CDF's. {To simulate the case where the PL $\bpi$ is not
  directly available, we generate one more independent reference sample
  path ${\bZ^{\scriptscriptstyle{( 0 )}}} = \{
  {Z_1^{\scriptscriptstyle{( 0 )}}, \ldots
    ,Z_{n_0}^{\scriptscriptstyle{( 0 )}}} \}$ of length $n_0 \gg
  \left|\bTheta\right| = N^2$, thus enabling us to obtain a good
  estimate of $\bpi$. Note that we do not rely on the test sample paths
  to estimate the PL $\bpi$.} The ground truth $\bpi$ is computed by
taking any row of $\bP^{m_0}$ for some sufficiently large $m_0$.

Having the ground truth PL $\bpi$ at hand, with {the test} sample paths
${\bZ^{\scriptscriptstyle{( t )}}} = \{ {Z_1^{\scriptscriptstyle{( t
      )}}, \ldots ,Z_n^{\scriptscriptstyle{( t )}}} \}$, $t = 1, \ldots,
T$, we can compute $T$ samples of the scalar random variable ${D}(
{{\boldsymbol{\Gamma} _n} \| {{{\bpi}}} } )$, by \eqref{2}. {Using
  these samples}, we obtain an empirical CDF of ${D}(
{{\boldsymbol{\Gamma} _n} \| {{{\bpi}}} } )$, denoted ${F}( \cdot; n )$,
which can be treated as a dependable proxy of the actual one. The
threshold given by \eqref{337} with $F_{\text{em}}(\cdot; n)$ replaced
by $F(\cdot; n)$ is then taken as a reliable proxy of
$\eta_{n,\beta}$. {We denote this proxy by $\eta^{*}_{n,\beta}$. To
  emphasize the dependence on $N$, we write $\eta_{n,\beta}$ (resp.,
  $\eta^{*}_{n,\beta}$) as $\eta_{n,\beta}(N)$ (resp.,
  $\eta^{*}_{n,\beta}(N)$).}  Next, using {the reference} sample path
${\bZ^{\scriptscriptstyle{( 0 )}}}$ and applying Alg.~\ref{alg:thres},
we obtain {$\eta_{n,\beta}^{\text{wc}}(N)$ and $\bar
  \eta_{n,\beta}^{\text{wc}}(N)$}.

Let the target false positive rate be $\beta = 0.001$. Consider four
different scenarios where $N$ is $2$, $4$, $6$, and $8$,
respectively. Set $\varepsilon = 10^{-10}$, $T = 1000$, $m_0 = 1000$,
and $n_0 = 1000N^2$. {Here we note that, in all our experiments, an
  estimate $\hat \bpi$ for $\bpi$ with $\|\hat \bpi - \bpi\| \le
  10^{-6}$ can be obtained by executing Alg. \ref{alg:thres} with
  parameters $n_0 \ge 500N^2$ and $\varepsilon \le 10^{-8}$.} In
Figs. \ref{fig1:N2} through \ref{fig1:N5}, the red line plots
$\eta^{*}_{n,\beta}(N)$, the blue line $\eta_{n,\beta}^{\text{wc}}(N)$,
the magenta line $\bar \eta_{n,\beta}^{\text{wc}}(N)$, and the green
line $\eta_{n,\beta}^{\text{sv}}(N)$ (cf. \eqref{335}), all as a
function of the sample size $n$. Setting sample sizes $n$ reasonably
small ($n$ should at least be comparable to $N^2$), it can be seen that
$\eta_{n,\beta}^{\text{wc}}(N)$ and $\bar \eta_{n,\beta}^{\text{wc}}(N)$
are more accurate than $\eta_{n,\beta}^{\text{sv}}$, except for the case
$N = 2$ where all estimators perform approximately equally well. In
particular, as $N$ increases, the estimation errors of
$\eta_{n,\beta}^{\text{wc}}(N)$ and $\bar \eta_{n,\beta}^{\text{wc}}(N)$
are consistently close to zero, while the approximation error of
$\eta_{n,\beta}^{\text{sv}}$ increases significantly. Moreover, for the
scenarios $N = 6, \,8$, $\eta_{n,\beta}^{\text{wc}}(N)$ and $\bar
\eta_{n,\beta}^{\text{wc}}(N)$ are very close.


\begin{figure*}[thpb]  
	\centering
	\begin{subfigure}[b]{0.495\textwidth}
		\includegraphics[width=\textwidth]{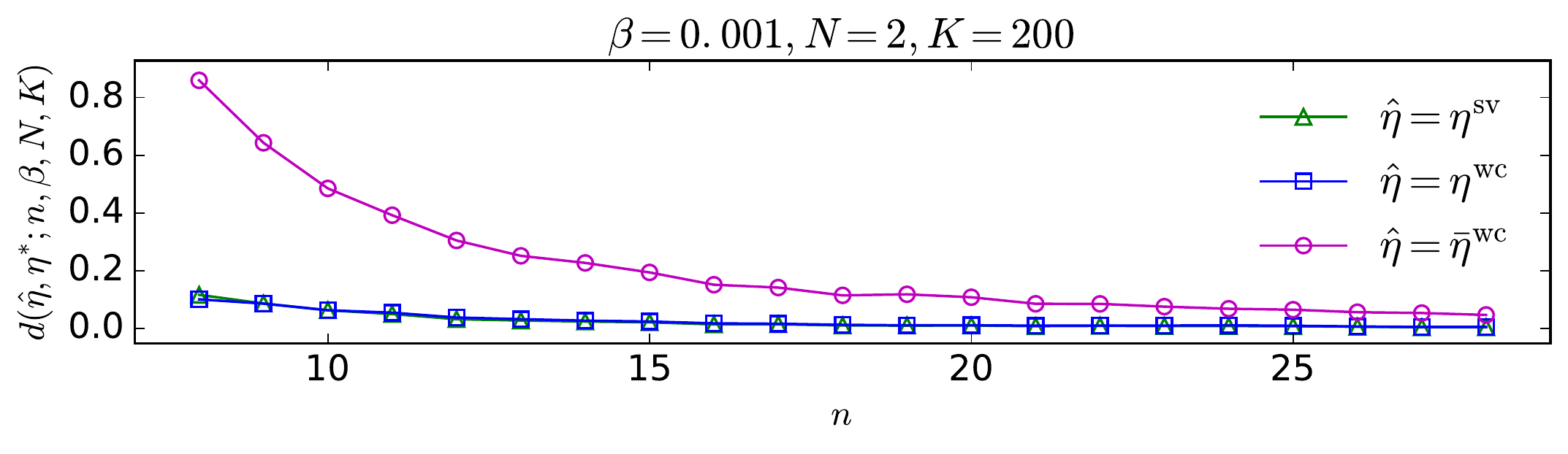} 
		\caption{}
		\label{fig1:errN2}
	\end{subfigure} 
	\begin{subfigure}[b]{0.495\textwidth}
		\includegraphics[width=\textwidth]{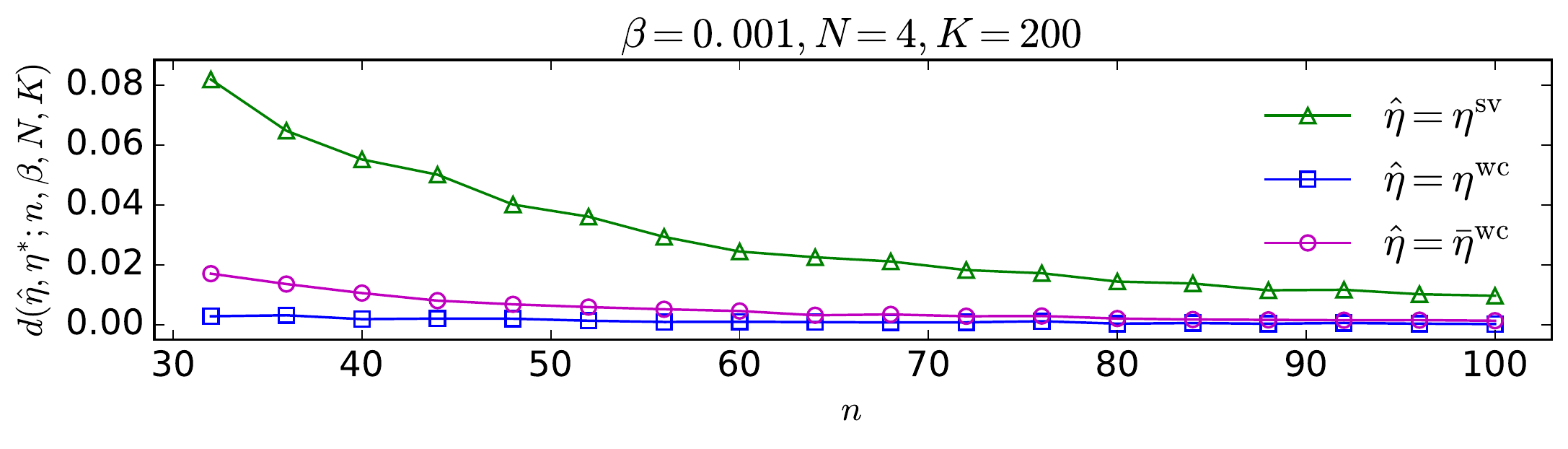}
		\caption{}
		\label{fig1:errN4}
	\end{subfigure}   
	\begin{subfigure}[b]{0.495\textwidth}
		\includegraphics[width=\textwidth]{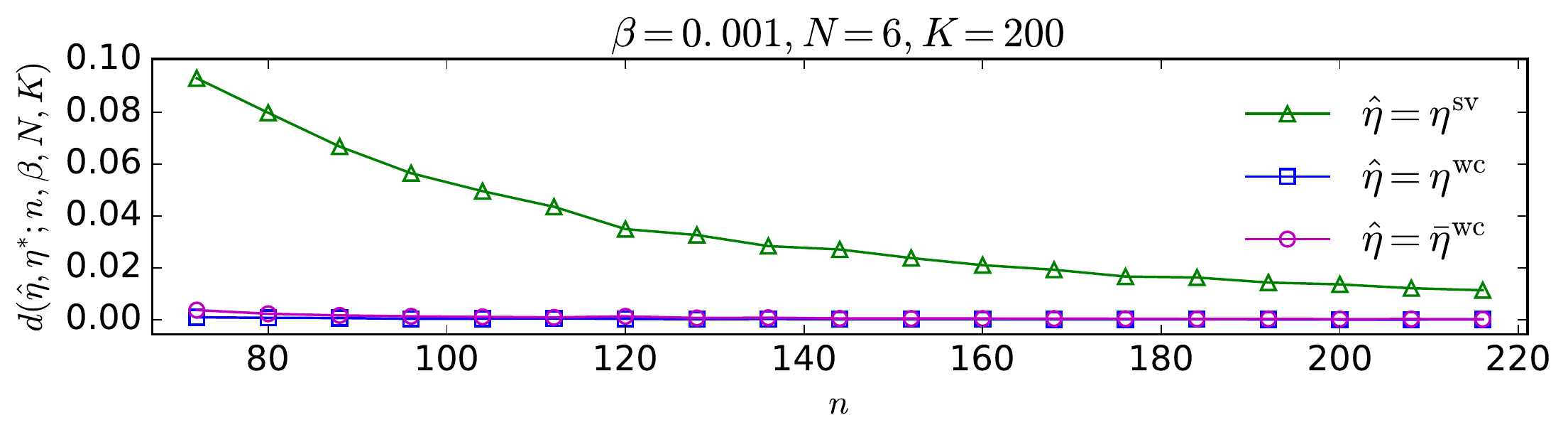}
		\caption{}
		\label{fig1:errN6}
	\end{subfigure}  
	\begin{subfigure}[b]{0.495\textwidth}
		\includegraphics[width=\textwidth]{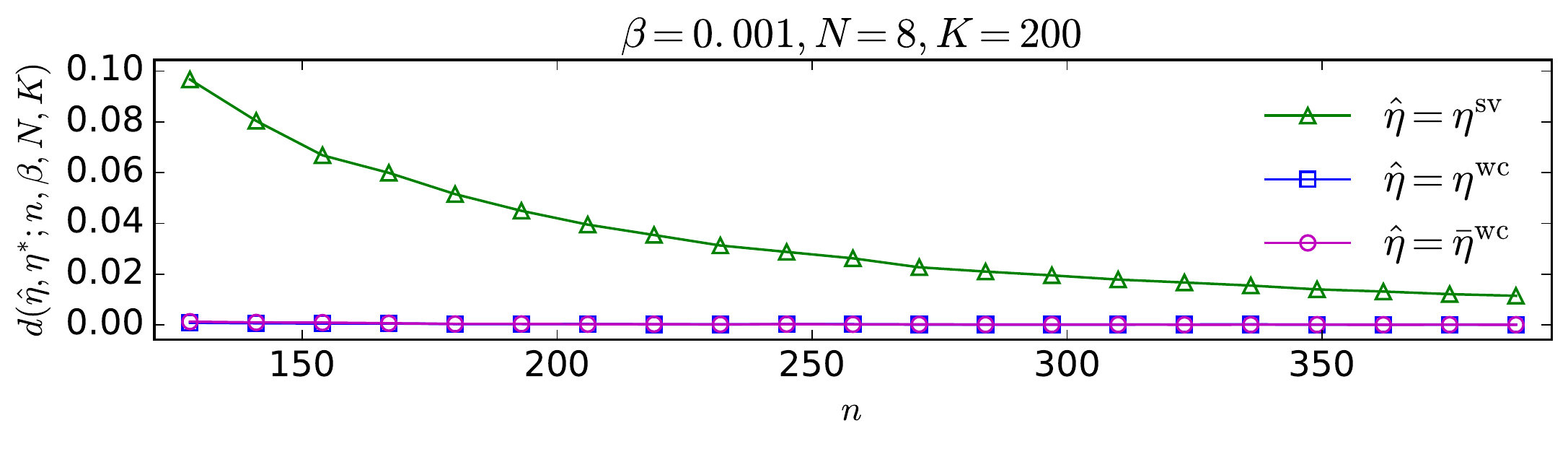}
		\caption{}
		\label{fig1:errN8}
	\end{subfigure}  
	\caption{Evaluation of average squared estimation errors for different types of threshold estimators.}
	\label{fig1:errN2468}
\end{figure*}


\begin{rem} \label{re5} \emph{In Figs. \ref{fig1:N2}-\ref{fig1:N5}, the red line representing the ``actual'' value
		$\eta^{*}_{n,\beta}$ is not smooth; this is because each time when varying the
		sample size $n$, we regenerate all the sample paths
		${{\bZ}^{\scriptscriptstyle{( t )}}} = \{ {Z_1^{\scriptscriptstyle{(
					t )}}, \ldots ,Z_n^{\scriptscriptstyle{( t )}}} \}, ~t = 1,
		\ldots ,T$ from scratch. On the other hand, the blue (resp., magenta) line
		corresponding to $\eta_{n,\beta}^{\text{wc}}$ (resp., $\bar \eta_{n,\beta}^{\text{wc}}$) is smooth because we only need to
		generate the $T$ Gaussian (resp., $\chi^2$-type) sample vectors once. In our experiments, most of the
		running time is spent generating the sample paths
		${{\bZ}^{\scriptscriptstyle{( t )}}}$ and calculating $\eta^{*}_{n,\beta}$
		therefrom. In practice, we
		will neither generate such samples nor calculate $\eta^{*}_{n,\beta}$, and only
		need to focus on obtaining $\eta_{n,\beta}^{\text{wc}}$ or $\bar \eta_{n,\beta}^{\text{wc}}$, which is 
		computationally not expensive.}
\end{rem}

\begin{rem} \label{re6} \emph{Theoretically speaking, we could use the
    ``actual'' threshold $\eta^{*}_{n,\beta}$ as obtained above, but it
    is of little practical value; the reason is that in statistical
    anomaly detection applications, we are typically faced with a long
    series of observations and want to use a so-called \emph{windowing
      technique} (see Sec. \ref{sec:simuNet}), which divides the
    observations into a sequence of detection windows with the same time
    length. The sample sizes $n$ in different windows may not
    necessarily be equal, leading to different threshold settings when
    sliding the windows. If we use the simulated ``actual'' threshold,
    then, when varying the detection windows, we will need to regenerate
    the corresponding samples (for threshold estimation purposes) from
    scratch, which is computationally too expensive, especially when
    there are many detection windows. In contrast, to compute our
    estimator $\eta_{n,\beta}^{\text{wc}}$ (resp., $\bar
    \eta_{n,\beta}^{\text{wc}}$), we only need to generate one set of
    Gaussian (resp., $\chi^2$-type) sample vectors (cf. Remark
    \ref{re5}), which can be shared by all the detection windows, thus,
    saving a lot of computation time. To see this more clearly, let us
    denote by $\tau_1$ the average running time for generating a set of
    samples with $T$ ($T = 1000$ is empirically a good choice) Gaussian
    (resp., $\chi^2$-type) vectors according to \eqref{336} (resp.,
    \eqref{chi2}), and $\tau_2$ the average running time for calculating
    a threshold via \eqref{337} given the corresponding sample vectors
    required to derive the empirical CDF. Clearly, we have $\tau_1 \gg
    \tau_2 > 0$. Assume we have $W$ detection windows. Then, if we
    directly simulate the statistic so as to estimate the threshold for
    each and every detection window, the total running time would be
    $c_1W\tau_1 + c_2W\tau_2 = (c_1\tau_1 + c_2\tau_2) W$, where $c_1,
    c_2 > 0$ are two scaling constants satisfying $c_1\tau_1 \gg
    c_2\tau_2$. On the other hand, by simulating Gaussian (resp.,
    $\chi^2$-type) samples, the total running time required to estimate
    all the thresholds for the $W$ detection windows would be $c_3\tau_1
    + c_4\tau_2W$, where $c_3, c_4 > 0$ are two scaling constants
    satisfying $c_4 \approx c_2$, leading to $0 < c_4\tau_2 \ll
    c_1\tau_1 + c_2\tau_2$. Thus, for large $W$ we have $c_3\tau_1 +
    c_4\tau_2W \ll (c_1\tau_1 + c_2\tau_2) W$.}
\end{rem}

{To further investigate the performance of different classes of
  threshold estimators, we now take the randomness of the transition
  matrix $\bP$ into account and define a simulation-based metric
  $d\left( {\hat \eta ,\eta^{*} ;n, \beta, N, K} \right)$ to quantify
  the average squared empirical estimation error, specified as follows:
  \begin{align}
	d\left( {\hat \eta ,\eta^{*} ;n, \beta, N, K} \right) = \frac{1}{{K}}\sum\limits_{k = 1}^{K} { { {\left( {\hat \eta^{\scriptscriptstyle{(k)}}_{n,\beta} \left( {N} \right) - \eta^{\scriptscriptstyle{*(k)}}_{n,\beta} \left( {N} \right)} \right)^2} } }.    \label{accuracy-metric}
\end{align}
Recall that $N$ is a parameter representing the number of states in the
original chain $\bY$. We denote by $\hat \eta$ the threshold estimator
class (could be $\eta^{\text{sv}}$, $\eta^{\text{wc}}$, or $\bar
\eta^{\text{wc}}$), and by $\eta^{*}$ a proxy of the actual threshold
class (derived by directly simulating the samples of the test
statistic). Denote by $K$ the number of independent repetitions of the
calculation for $( {\hat \eta^{\scriptscriptstyle{(k)}}_{n,\beta} ( {N}
  ) - \eta^{\scriptscriptstyle{*(k)}}_{n,\beta} ( {N} )} )^2$, where
$\hat \eta^{\scriptscriptstyle{(k)}}_{n,\beta} ( {N} )$
(resp., $\eta^{\scriptscriptstyle{*(k)}}_{n,\beta} ( {N} )$)
denotes the class $\hat \eta$ (resp., $\eta^{*}$) instantiated under
parameters $n$, $\beta$, $N$, and $k \in \{1, \ldots, K\}$.}

{Setting $\beta = 0.001$, $K = 200$, $N \in \{2, \,4, \,6, \,8\}$,
  and $n \in \left\{ {\bar n = 2{N^2} + i \times \left\lfloor {0.2{N^2}
          + 1} \right\rfloor : \bar n < 6{N^2} + 5,i \in \mathbb{N_+}}
  \right\}$, we evaluate $d\left( {\hat \eta ,\eta^{*} ;n, \beta, N, K}
  \right)$. The results are shown in Fig. \ref{fig1:errN2468}. Several
  observations can be made from Figs. \ref{fig1:errN2}-\ref{fig1:errN8}:
  (\rmnum{1}) Except for the case $N = 2$, both $\eta^{\text{wc}}$ and
  $\bar \eta^{\text{wc}}$ outperform $\eta^{\text{sv}}$, that is,
  $d\left( {\eta^{\text{wc}} ,\eta^{*} ;n, \beta, N, K} \right) <
  d\left( {\eta^{\text{sv}} ,\eta^{*} ;n, \beta, N, K} \right)$ and
  $d\left( {\bar \eta^{\text{wc}} ,\eta^{*} ;n, \beta, N, K} \right) <
  d\left( {\eta^{\text{sv}} ,\eta^{*} ;n, \beta, N, K}
  \right)$. (\rmnum{2}) For the cases $N = 6, \,8$, $\eta^{\text{wc}}$
  and $\bar \eta^{\text{wc}}$ perform almost equally well, with both
  $d\left( {\eta^{\text{wc}} ,\eta^{*} ;n, \beta, N, K} \right)$ and
  $d\left( {\bar \eta^{\text{wc}} ,\eta^{*} ;n, \beta, N, K} \right)$
  being very close to zero and, for the cases $N = 2, \,4$,
  $\eta^{\text{wc}}$ outperforms $\bar \eta^{\text{wc}}$, i.e., 
  $d\left( { \eta^{\text{wc}} ,\eta^{*} ;n, \beta, N, K} \right) <
  d\left( {\bar \eta^{\text{wc}} ,\eta^{*} ;n, \beta, N, K}
  \right)$. (\rmnum{3}) Only for the case $N = 2$, $\eta^{\text{sv}}$
  performs the best among the three estimators and, $\eta^{\text{wc}}$
  performs approximately equally well with $\eta^{\text{sv}}$ in this
  case. More extensive comparison results can be derived using
  TAM~\cite{TAHTMA}. We 
  may empirically conclude that, $\eta^{\text{wc}}$ performs
  consistently the best among the three for almost all scenarios that we
  have considered and, on the other hand, $\eta^{\text{sv}}$ performs
  unsatisfactorily when $N > 2$, and $\bar \eta^{\text{wc}}$ is
  numerically not as stable as $\eta^{\text{wc}}$, especially for the
  cases where $N \le
  4$.} 

\begin{figure*}[ht]  
	\centering
	\begin{subfigure}[b]{0.495\textwidth}
		\includegraphics[width=\textwidth]{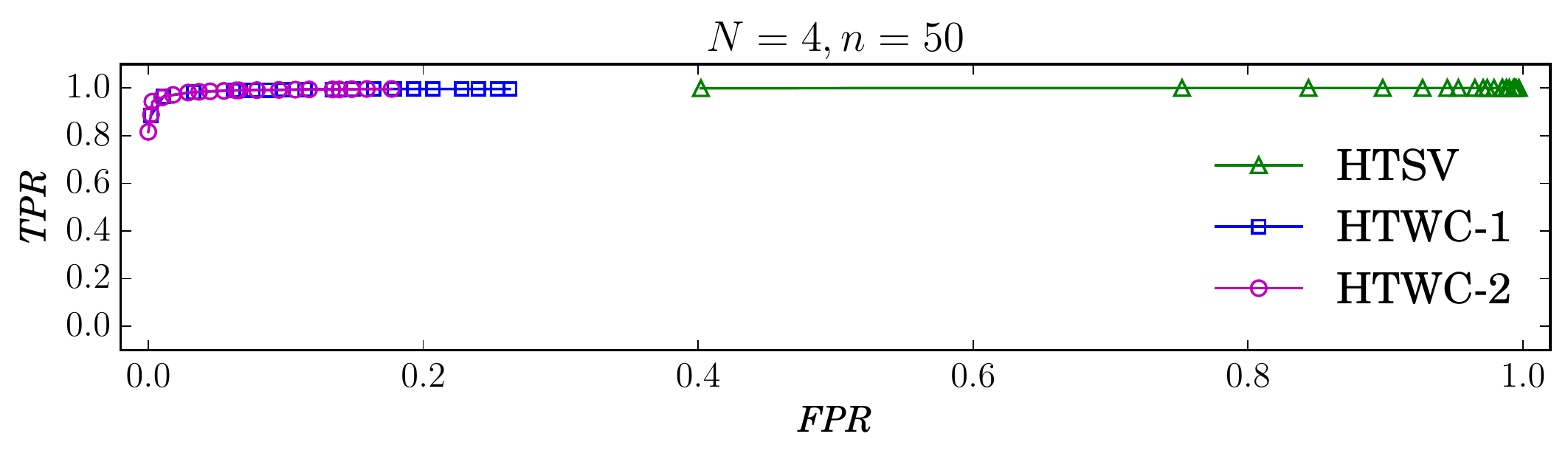}
		\caption{}
		\label{ROC-gr4}
	\end{subfigure}  
	\begin{subfigure}[b]{0.495\textwidth}
		\includegraphics[width=\textwidth]{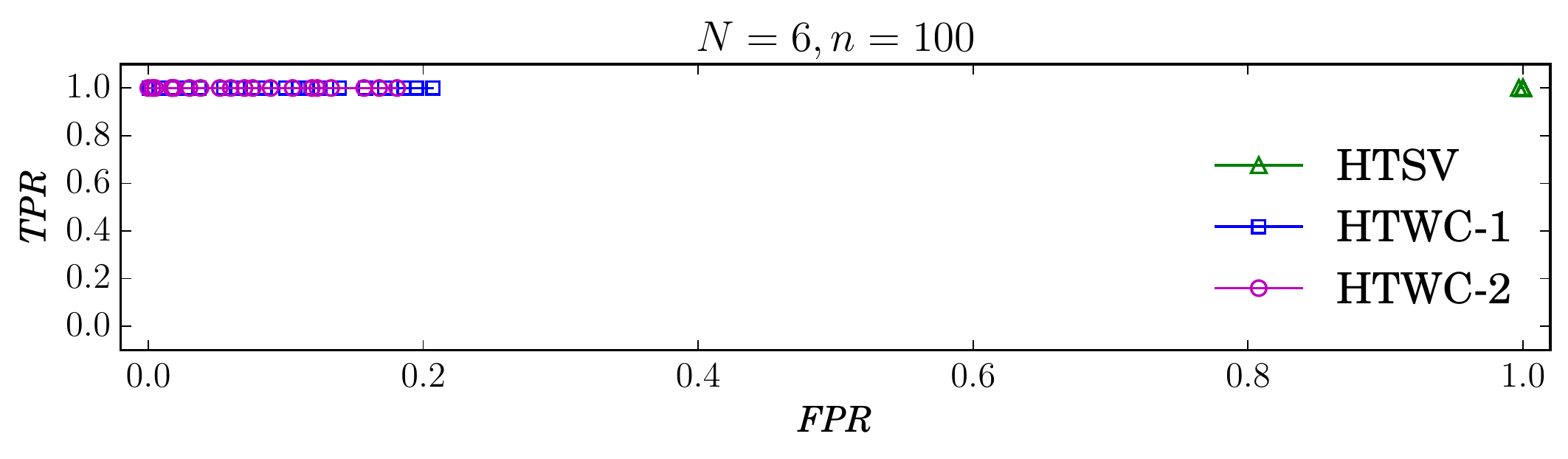}
		\caption{}
		\label{ROC-gr6}
	\end{subfigure} 
	\caption{Results from ROC analysis of the ordinary Hoeffding test.}
	\label{fig:ROCana}
\end{figure*}

\subsection{ROC Analysis for the Hoeffding Test with 
Different Threshold Estimators} \label{sec:ROCHoeff}

In this subsection, for simplicity and economy of space, we again only consider the ordinary (and not the robust) Hoeffding test. We note here that similar results can be derived for the robust Hoeffding test. The numerical experiments are conducted using the software package ROM \cite{ROCHM}.

Let ${\bTheta} = \{1,2,\ldots,N^2\}$ containing $N^2$
states. For a given sample size $n$ and a given target false positive rate (FPR) $\beta$, the
three thresholds $\eta_{n,\beta}^{\text{wc}}$, $\bar \eta_{n,\beta}^{\text{wc}}$, and $\eta_{n,\beta}^{\text{sv}}$, respectively, give rise
to three different \emph{discrete tests} (denote them by ``HTWC-1,'' ``HTWC-2,'' and ``HTSV,''
respectively). To compare their performances, we will conduct the Receiver
Operating Characteristic (ROC) \cite{fawcett2006introduction} analysis (detection rate vs. false alarm
rate) using simulated data.


\begin{table}
	\centering
	\caption{ROC points vs. target FPR ($N = 4$, $n = 50$).} \label{tab-roc}
	\begin{tabular}{rrrrrrr}
		\toprule[1.2pt]
		\multirow{2}{*}{target FPR $\beta$} &
		\multicolumn{2}{c}{HTWC-1} &
		\multicolumn{2}{c}{HTWC-2} &
		\multicolumn{2}{c}{HTSV} \\
		& {FPR} & {TPR}& {FPR} & {TPR} & {FPR} & {TPR} \\
		\midrule[1.2pt]
		0.001 & 0.002 & 0.885 & 0.0 & 0.816 & 0.402 & 0.999 \\
		0.01 & 0.011 & 0.965 & 0.002 & 0.888 & 0.752 & 1.0 \\
		0.02 & 0.018 & 0.983 & 0.003 & 0.943 & 0.844 & 1.0 \\
		0.03 & 0.025 & 0.99 & 0.01 & 0.96 & 0.898 & 1.0 \\
		0.04 & 0.038 & 0.99 & 0.018 & 0.971 & 0.927 & 1.0 \\
		0.05 & 0.047 & 0.991 & 0.029 & 0.981 & 0.945 & 1.0 \\
		\bottomrule[1.2pt]
	\end{tabular}
\end{table}

\begin{table}
	\centering
	\caption{ROC points vs. target FPR ($N = 6$, $n = 100$).} \label{tab-roc-2}
	\begin{tabular}{rrrrrrr}
		\toprule[1.2pt]
		\multirow{2}{*}{target FPR $\beta$} &
		\multicolumn{2}{c}{HTWC-1} &
		\multicolumn{2}{c}{HTWC-2} &
		\multicolumn{2}{c}{HTSV} \\
		& {FPR} & {TPR}& {FPR} & {TPR} & {FPR} & {TPR} \\
		\midrule[1.2pt]
		0.001 & 0.001 & 1.0 & 0.0 & 1.0 & 0.997 & 1.0 \\
		0.01 & 0.008 & 1.0 & 0.003 & 1.0 & 1.0 & 1.0 \\
		0.02 & 0.017 & 1.0 & 0.005 & 1.0 & 1.0 & 1.0 \\
		0.03 & 0.028 & 1.0 & 0.017 & 1.0 & 1.0 & 1.0 \\
		0.04 & 0.037 & 1.0 & 0.017 & 1.0 & 1.0 & 1.0 \\
		0.05 & 0.055 & 1.0 & 0.019 & 1.0 & 1.0 & 1.0 \\
		\bottomrule[1.2pt]
	\end{tabular}
\end{table}


{Similar to what we have done in Sec. \ref{sec:num.A}, we first randomly
create a valid $N \times N$ transition matrix $\bQ$, hence an $N^2
\times N^2$ transition matrix $\bP$, and then generate $T$ sample paths
of the chain $\bZ$, each with length $n$, denoted
${\bZ^{\scriptscriptstyle{( t )}}} = \{ {Z_1^{\scriptscriptstyle{( t
      )}}, \ldots ,Z_n^{\scriptscriptstyle{( t )}}} \}$, $t = 1, \ldots
,T$. {From $\bP$ we derive the PL $\bpi$}. Next, to simulate anomalies, we create another valid $N \times N$
transition matrix $\bar {\bQ}$, hence an $N^2 \times N^2$ transition
matrix $\bar {\bP}$, and generate $T$ sample paths of the corresponding
chain $\bar {\bZ}$, each with length $n$, denoted ${{\bar
    {\bZ}}^{\scriptscriptstyle{( t )}}} = \{ {{\bar
    Z}_1^{\scriptscriptstyle{( t )}}, \ldots ,{\bar
    Z}_n^{\scriptscriptstyle{( t )}}} \}$, $t = 1, \ldots ,T$. Label
each sample path of $\bar {\bZ}$ (resp., ${\bZ}$) with length $n$ as
``positive'' (resp., ``negative''). Then, $\{{\bZ^{\scriptscriptstyle{(
      t )}}}:\,t \in \{1,
\ldots ,T\}\} \cup \{{\bar {\bZ}}^{\scriptscriptstyle{( t )}}:\,t \in \{1,
\ldots ,T\}\}$ will be our test set, which contains $T$ negative ($\bZ^{\scriptscriptstyle{( t )}}$) and $T$ positive
(${\bar {\bZ}}^{\scriptscriptstyle{( t )}}$) sample paths.}

{Now, by executing Alg. \ref{alg:thres} without estimating $\bpi$
  (since the ground truth is available), we obtain
  $\eta_{n,\beta}^{\text{wc}}$ and $\bar
  \eta_{n,\beta}^{\text{wc}}$. Also, by \eqref{335} we obtain
  $\eta_{n,\beta}^{\text{sv}}$. For each sample path in the test set, we
  compute ${D}( {{\boldsymbol{\Gamma} _n}\| {{{\bpi}}} } )$ by
  \eqref{2}. Next, using $\eta_{n,\beta}^{\text{wc}}$ (resp., $\bar
  \eta_{n,\beta}^{\text{wc}}$, $\eta_{n,\beta}^{\text{sv}}$), we can
  apply HTWC-1 (resp., HTWC-2, HTSV) to detect each sample path as
  positive or negative. Then, we integrate these reports with the ground
  truth labels so as to calculate the true positive rate (TPR) and FPR,
  thereby, obtaining a point of the ROC space.}

{In our experiments, we take $T = 1000$. Fig. \ref{ROC-gr4} (resp.,
  \ref{ROC-gr6}) shows the ROC graphs of HTWC-1, HTWC-2, and HTSV for a
  scenario corresponding to $N = 4,\, n = 50$ (resp., $N = 6,\, n =
  100$); different points on the graph are obtained by $\beta$ taking
  values from a predesignated finite set $\{0.001\} \cup \{0.01, 0.02,
  \ldots, 0.19\}$.  It is seen from Fig. \ref{ROC-gr4} (or
  Fig. \ref{ROC-gr6}) that all TPR values are very close to 1, which is
  good, but for most cases (each case corresponds to a specific
  ``small'' target FPR $\beta$) HTWC-1 and HTWC-2 have much closer FPR
  values to the target FPR value than HTSV, meaning HTWC-1 and HTWC-2
  are able to control for false alarms better than HTSV. To see this
  more clearly, we show a few specific values of the (TPR, FPR) pair in
  Tables~\ref{tab-roc} and \ref{tab-roc-2}. It is worth noting that in the $N = 6$
  scenario, HTSV is almost a random guess for all the target FPR cases
  that are considered.  More extensive experiments show that, as $N$
  increases, the performance of HTSV gets worse and worse; in
  particular, when $N \ge 6$, HTSV is very likely merely a random guess
  yielding an ROC point close to $(1,1)$. During our experiments,
  another observation is that, for each fixed $N$ and $\beta$, when $n$
  increases, all HTWC-1, HTWC-2, and HTSV perform better and better;
  this is because with larger sample sizes, all the three estimators
  $\eta_{n,\beta}^{\text{wc}}$, $\bar \eta_{n,\beta}^{\text{wc}}$, and
  $\eta_{n,\beta}^{\text{sv}}$ approximate the actual $\eta_{n,\beta}$
  better. We therefore conclude that HTWC-1 (or HTWC-2) typically
  outperforms HTSV in the sense that the former has a better capability
  of controlling the false alarm rate (i.e., FPR) while maintaining a
  satisfactory detection rate (i.e., TPR).}

{
  \begin{rem} \label{remark-roc} \emph{A natural concern about the ROC
      analysis above might be the setting of the target FPR ($\beta$)
      values; one may ask: How about always setting $\beta$ to a ``very
      small'' value, say, $10^{-10}$, $10^{-100}$, or even $10^{-1000}$?
      We have actually already discussed this partly in
      Sec. \ref{sec:prob}. Setting a too small $\beta$ would typically
      lead to an unsatisfactory detection rate (TPR). In addition, note
      that $\eta_{n,\beta}^{\text{wc}}$ (or $\bar
      \eta_{n,\beta}^{\text{wc}}$) is numerically obtained from an
      empirical CDF, say, $G(x)$, of some scalar random variable; we
      have $G(x)$ nondecreasing, and $\mathop {\lim }\nolimits_{x \to +
        \infty } G\left( x \right) = 1$, implying that finding an
      ``accurate'' $x$ such that $G(x) = 1 - \beta$ would be hard for a
      too small $\beta \in (0 ,1)$. An empirically ``good'' choice of
      $\beta$ is 0.001 (see Tabs. \ref{tab-roc} and \ref{tab-roc-2}),
      which is what we use in our applications. Because HTWC-1 and
      HTWC-2 perform almost equally well in our experiments, but HTWC-1
      is more stable and less computationally demanding, we will
      only apply HTWC-1 in the following.}
\end{rem}}

\subsection{Simulation Results for Network Anomaly Detection} \label{sec:simuNet}

\begin{figure}[thpb]
   \centering
   \includegraphics[scale=0.28]{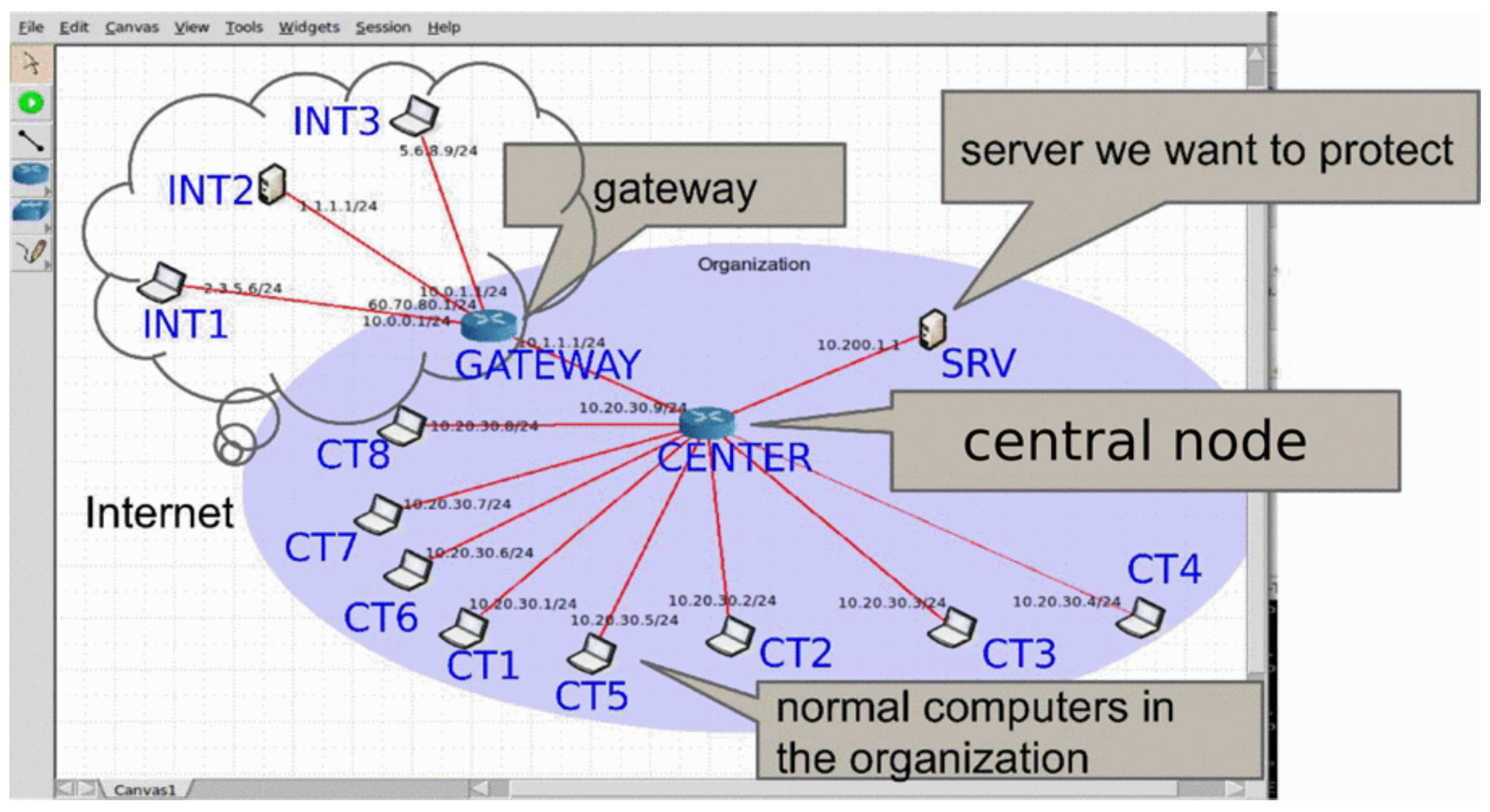}
   \caption{Simulation setting (from \cite{robust-anomaly-tcns}).}
\label{simulation_setting}
\end{figure}

\begin{figure*}[t]  
	\centering
	\begin{subfigure}[b]{0.495\textwidth}
		\includegraphics[width=\textwidth]{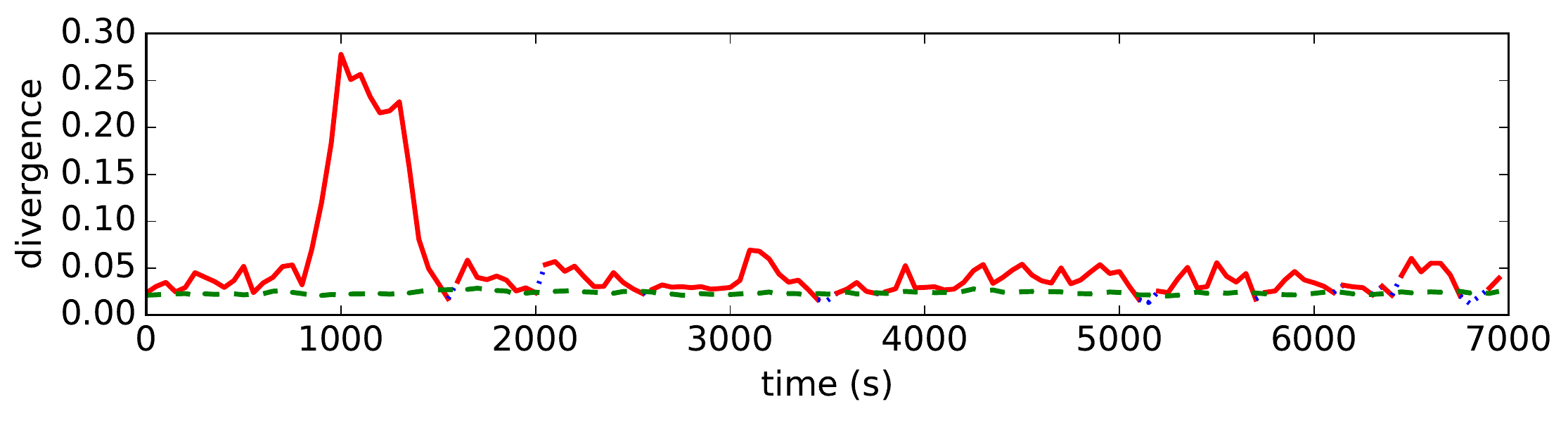}
		\caption{Sanov}
		\label{LargeFileSanovN8}
	\end{subfigure} 
	\begin{subfigure}[b]{0.495\textwidth}
		\includegraphics[width=\textwidth]{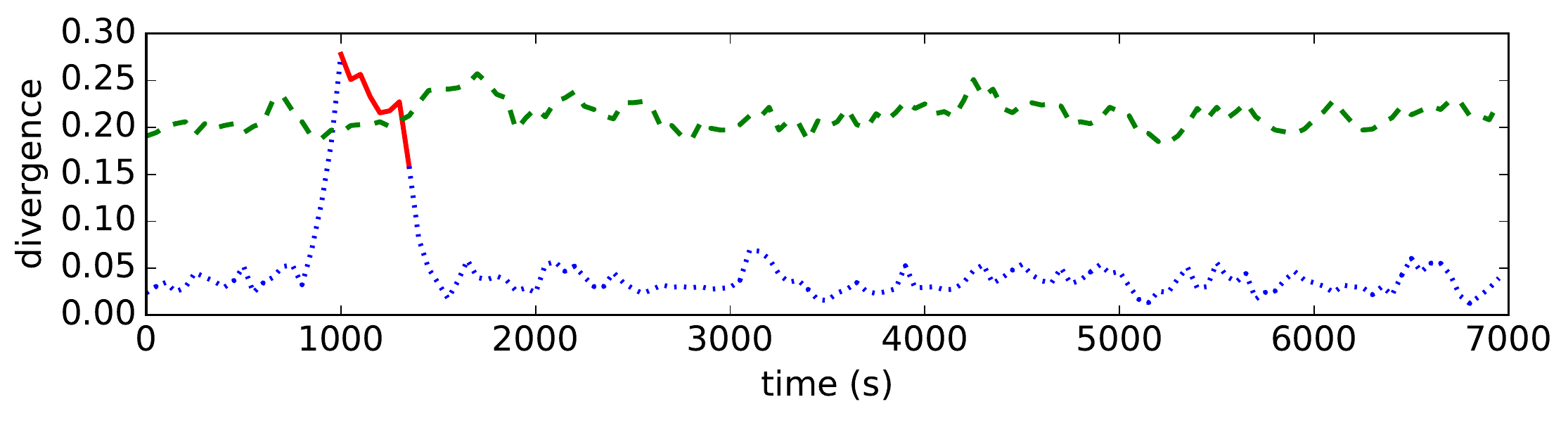}
		\caption{WC}
		\label{LargeFileTaylorN8}
	\end{subfigure}   
	\caption{Detection results for Scenario \ref{sec:simuNet}-1 with
		$w_{\text{d}} = 50$ s, $w_{\text{s}} = 200$ s, $k =2$, $n_1 = 1$, $n_2 = 2$,
		$n_3 = 2$; (a) threshold is estimated by use of Sanov's
		theorem; (b) threshold is estimated by use of the weak convergence
		result.}
\end{figure*}

In this subsection we test our approach in a communication network
traffic anomaly detection application. We will use the term
\textit{traffic} and \textit{flow} interchangeably. We perform the
simulations using the software package SADIT \cite{SADIT}, which, based
on the \textit{fs}-simulator \cite{fs}, is capable of efficiently
generating flow-level network traffic datasets with annotated anomalies.

As shown in Fig. \ref{simulation_setting}, the simulated network
consists of an internal network involving eight normal users
(\textit{CT1-CT8}), a server (\textit{SRV}) that stores sensitive
information, and three Internet nodes (\textit{INT1-INT3}) that connect
to the internal network via a gateway (\textit{GATEWAY}).

As in \cite[Sec. III.A]{robust-anomaly-tcns}, to characterize the
statistical properties of the flow data, we use as features the flow
duration and size (bits). We also cluster the source/destination IP
addresses and use as features for each flow the assigned cluster ID and
the distance of the flow's IP from the cluster center. For each feature,
we quantize its values into discrete symbols so as to obtain a finite
alphabet $\boldsymbol{\Xi}$, hence ${\bTheta}$, for our
model. Based on the time stamps (the start times) of the flows, we
divide the flow data into a series of detection windows, each of which
contains a set of flow observations (see \cite{robust-anomaly-tcns} for
details).

To implement our anomaly detection approach, we first estimate a PL
$\bpi$ (resp., a PL set $\boldsymbol{\Pi}$) from the stationary (resp.,
time-varying) normal traffic. Note that, for either case, the reference
data should be anomaly-free ideally. However, in our experiments, for
the stationary case we use as reference traffic the entire flow
sequence with anomalies injected at some time interval; this makes sense
because the size of a typical detection window is much smaller than that
of the whole flow sequence and the fraction of anomalies is indeed very
small, leading to an estimation for the PL with acceptable accuracy. On
the other hand, for the time-varying case we generate the reference
traffic without anomalies and the test traffic with anomalies
separately, sharing all the parameter settings in the statistical model
used in SADIT except the ones for introducing anomalies.  Note that,
estimating a PL for the stationary traffic is relatively easy, while,
for the time-varying traffic, we need to make an effort to estimate
several different PLs corresponding to certain periods of the day. We
apply the two-step procedure proposed in \cite{robust-anomaly-tcns};
that is, we first generate a relatively large PL set and then refine the
candidate PLs therein by solving a \textit{weighted set cover
  problem}. Note also that, if we already know the periodic system
activity pattern, then we can directly estimate the PL set period by
period; see another anomaly detection application in
Sec.~\ref{sec:anoWaze} for example.

Now, having the reference PL (resp., PL set) at hand, we persistently
monitor the test traffic and report an anomaly instantly as long as the
relative entropy $D( { {{\boldsymbol{\Gamma} _n}} \|{\bpi} } )$ (resp.,
$\mathop {\inf }_{\bpi \in \boldsymbol{\Pi} } D( {{\boldsymbol{\Gamma}
    _n} \| \bpi } )$) exceeds the threshold $\eta_{n,\beta}^{\text{wc}}$ for the
current detection window, where $n$ is the number of flow samples within
the window. It is worth pointing out that, for the current application,
we will not seek to identify which flows belonging to an abnormal
detection window contribute mostly to causing the anomaly, but, in some
other applications, e.g., the one in Sec.~\ref{sec:anoWaze}, we will do
so.

In the following, we consider two scenarios -- one for stationary
traffic and the other for time-varying traffic.

\subsubsection{Stationary Network Traffic -- Scenario
  \ref{sec:simuNet}-1} 

We mimic anomalies caused by a large file download
\cite[Sec. IV.A.2]{wa-ro-ca-pa-anomaly-cdc13}. The simulation time is
$7000$ s. A user increases its mean flow size to $10$ times the usual
value between $1000$ s and $1500$ s. The interval between the starting
points of two consecutive time windows is taken as $w_{\text{d}} = 50$ s, the
window-size is set to $w_{\text{s}} = 200$ s, and the target false positive rate
is set to $\beta = 0.001$. The number of user clusters is $k = 2$ and
the quantization level for flow duration, flow size, and distance to
cluster center is set to $n_1 = 1$, $n_2 = 2$, and $n_3 = 2$,
respectively. Thus, the original chain has $N = 2 \times 1 \times 2
\times 2 = 8$ states, and we have $N^2 = 64$ states in the transformed
chain.

The detection results are shown in Figs. \ref{LargeFileSanovN8} and
\ref{LargeFileTaylorN8}, both of which depict the relative entropy
(divergence) metric defined in \eqref{2}. The green dashed line in
Fig.~\ref{LargeFileSanovN8} is the threshold estimated using Sanov's
theorem (i.e., $\eta_{n,\beta}^{\text{sv}}$ given by \eqref{335}, where $n$ is the
sample size in each specific detection window). The green dashed line in
Fig. \ref{LargeFileTaylorN8} is the threshold given by our estimator
(i.e., $\eta_{n,\beta}^{\text{wc}}$ computed by Alg. \ref{alg:thres}). The interval
during which the divergence curve is above the threshold line (the red
segment) corresponds to the time instances reported as abnormal.
Fig. \ref{LargeFileSanovN8} shows that, if $\eta_{n,\beta}^{\text{sv}}$ is used as the
threshold, then the Hoeffding test reports too many false alarms, and,
Fig. \ref{LargeFileTaylorN8} shows that, if, instead, we use
$\eta_{n,\beta}^{\text{wc}}$ as the threshold, then the Hoeffding test does not report
any false alarm while successfully identifying the true anomalies
between $1000$ s and $1500$ s.

\subsubsection{Time-Varying Network Traffic -- Scenario
  \ref{sec:simuNet}-2}

Consider the case where the network in Fig. \ref{simulation_setting} is
simulated with a day-night traffic pattern in which the flow size
follows a log-normal distribution. We use precisely the same scenario as
that in \cite[Sec. IV.B.2]{robust-anomaly-tcns}. The ground truth
anomaly (consider an anomaly where node \textit{CT2} increases its mean
flow size by 30\%) is injected beginning at $59$ h and lasting for $80$
minutes.

\begin{figure}[ht]  
	\centering
	\includegraphics[width=0.485\textwidth]{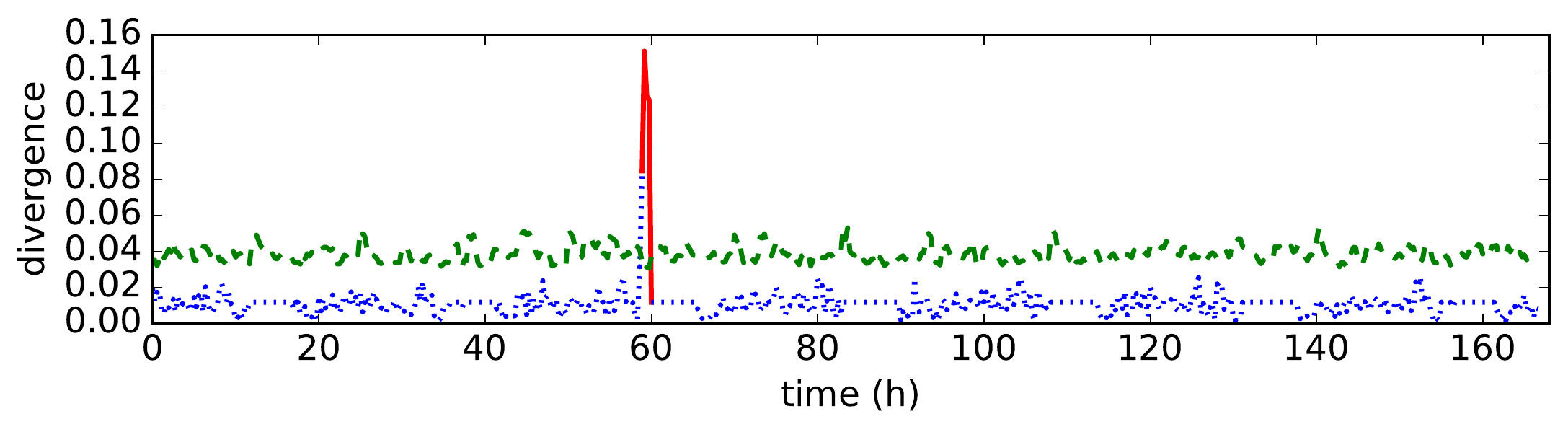}
	\caption{Detection result for Scenario \ref{sec:simuNet}-2 with $w_{\text{d}} = 1000$ s, $w_{\text{s}} = 1000$ s, $k =1$, $n_1 = 1$, $n_2 = 4$, $n_3 = 1$.}
	\label{time-varying-detect}
\end{figure}
Using the two-step procedure proposed in
\cite[Sec. III.C]{robust-anomaly-tcns}, we first obtain $32$ rough PL
candidates. Then, using the PL refinement algorithm given in
\cite[Sec. III.D]{robust-anomaly-tcns} equipped with the cross-entropy
threshold parameter $\lambda = 0.028$, which is
determined by applying Alg. \ref{alg:thresrob1}, we finally obtain $6$
PLs, being active during \textit{morning}, \textit{afternoon},
\textit{evening}, \textit{mid-night}, \textit{dawn}, and \textit{the
  transition time around sunrise}, respectively. Note that, since we
have obtained the PL set in a different way, in the following, when
applying Alg. \ref{alg:thresrob1} for each detection window, we can skip
the first two steps (lines \ref{line2} and \ref{line3}). In the
subsequent detection procedure, the chief difference between our method
and the one used in \cite{robust-anomaly-tcns} is that we no longer set
the threshold universally as a constant; instead, we calculate the
threshold $\eta_{n,\beta}^{\text{wc}}$ for each detection window using
Alg. \ref{alg:thresrob1}. Set $k = 1$, $n_1 = 1$, $n_2 = 4$, and $n_3 =
1$. Thus, the original chain has $N = 1 \times 1 \times 4 \times 1 = 4$
states, and we have $N^2 = 16$ states in the transformed chain for this
case. Take $w_{\text{d}} = 1000$ s, $w_{\text{s}} = 1000$ s, and $\beta = 0.001$.  We see
from Fig. \ref{time-varying-detect} that the anomaly is successfully
detected, without any false alarms.

\subsection{Anomaly Detection for Waze Jams} \label{sec:anoWaze}

\subsubsection{Dataset Description}

The Waze datasets under investigation are kindly provided to us by the
Department of Innovation and Technology (DoIT) in the City of
Boston. The datasets include three parts: the jam data $\scrJ_1$
(traffic slowdown information generated by Waze based on users' location
and speed; note that each jam consists of a set of points), the
corresponding point data $\scrJ_2$ (latitudes and longitudes of the
points within jams), and the alert data $\scrJ_3$ (traffic incidents
reported by active users; we will call such a user a ``Wazer''). For
each part, we only list the features that we have used in our
algorithms. In particular, each entry (jam) in $\scrJ_1$ has the
following fields: uuid (unique jam ID), start time, end time, speed
(current average speed on jammed segments in meters per second), delay
(delay caused by the jam compared to free flow speed, in seconds), and
length (jam length in meters). The information for each entry in
$\scrJ_2$ includes a jam uuid and the locations (latitudes and
longitudes) of the points within the jam. The fields of each entry in
$\scrJ_3$ include: uuid (unique system ID; this is different from the
jam ID in $\scrJ_1$), location (latitude and longitude per report), type
(event type; e.g., accident, weather hazard, road closed, etc.), start
time, and end time. It is seen that, by combing $\scrJ_1$ and $\scrJ_2$,
we can denote each jam in $\scrJ_1$ as
\[ 
(i, \text{uuid}[i], \text{loc}[i], \text{speed}[i], \text{delay}[i],
\text{length}[i], \text{startTime}[i]),
\]
where $i$ is the index, uuid is the unique jam ID, ``loc'' (resp.,
``startTime'') is the abbreviation for location (resp., start
time). Because we are only interested in detecting the abnormal jams in
real-time, we will not use the jam end times.

\subsubsection{Anomaly Description}

Typically we can observe lots of jams in certain areas during rush hour,
e.g., the AM/PM peaks, and most of them are ``normal'' except those with
extremely atypical features (delay, length, etc.). On the other hand, if
a jam was observed outside of rush hours or typical areas, then it would
likely be ``abnormal.''

\subsubsection{Description of the Experiments}

\begin{figure}[t]
	\centering
	\includegraphics[height=4.15cm]{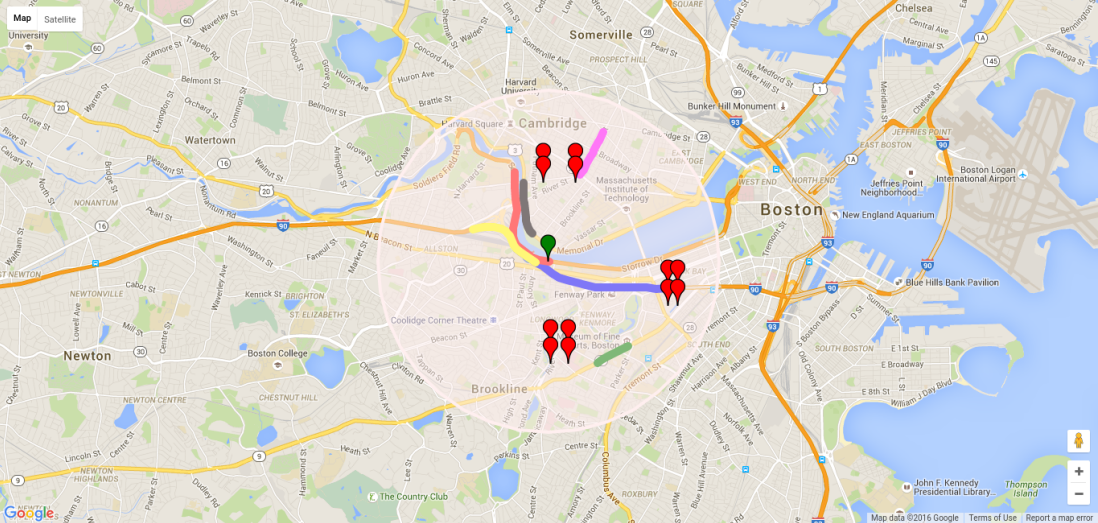}
	\caption{Location cluster centers and detected abnormal jams for
          a circle area around Boston University.}
	\label{fig:locClus}
\end{figure}

Treating Waze jams as a counterpart of the network flows in
Sec. \ref{sec:simuNet}, we implement the robust Hoeffding test on the
quantized jam data in the following experiments.

Consider an area around the Boston University (BU) bridge, whose
location is specified by latitude and longitude $(42.351848,
-71.110730)$ (see the green marker in Fig. \ref{fig:locClus}). Extract
the jam data no farther than 3 kilometers from BU (within the circle in
Fig. \ref{fig:locClus}). Note that it is possible for Waze to report
several jams at the same time. To assign each jam a unique time stamp,
we slightly perturb the start time of the jams that share the same time
stamp in the raw data. Such slight adjustments would not alter the
original data significantly.

Reference (resp., test) data are taken as jams reported on March 9, 2016
(resp., March 16, 2016). Both dates are Wednesdays, representing typical
workdays. There are 3218 jams in the reference data, and 3882 jams in
the test data.  Note that we have historical data for a relatively long
time period (compared to the test data within a detection window);
including all the jams reported within the selected reference time
period would not hurt the accuracy of the PLs (anomaly-free ideally) to
be estimated.

The features that we use for anomaly detection are location, speed,
delay, and length. The time stamp of a jam is taken as its start time.
To quantize the location, we need to define the distance between two
jams. For any valid index $i$, denote the complete location data of jam
$i$ by
\begin{equation}
\widehat{\text{loc}}[i] = \{(x_{i,1}, y_{i,1}), \ldots, (x_{i,i_n},
y_{i,i_n})\}, \label{locC} 
\end{equation}
where $x$'s and $y$'s denote the latitudes and longitudes, respectively,
and $i_n$ is the number of points in jam $i$ (typically, $i_n$ is
greater than $4$). Noting that most of the jams are approximately linear
in shape, we simplify \eqref{locC} by using the 4 vertices of the
``smallest'' rectangle that covers all the points in the jam and 
update \eqref{locC} by
\begin{align}
\text{loc}[i] = \{&(x_{i,\text{min}}, y_{i,\text{min}}), (x_{i,\text{min}}, y_{i,\text{max}}), \notag \\ &(x_{i,\text{max}}, y_{i,\text{min}}), (x_{i,\text{max}}, y_{i,\text{max}})\}, \label{locS}
\end{align}
where $ {x_{i,\min }} = \min \{ {{x_{i,1}}, \ldots ,{x_{i,{i_n}}}} \}$,
${x_{i,\max }} = \max \{ {{x_{i,1}}, \ldots ,{x_{i,{i_n}}}} \}$,
${y_{i,\min }} = \min \{ {{y_{i,1}}, \ldots ,{y_{i,{i_n}}}} \}$, and
${y_{i,\max }} = \max \{ {{y_{i,1}}, \ldots ,{y_{i,{i_n}}}} \} $.  Note
that $\text{loc}[i]$ in \eqref{locS} only contains $4$ points.  Denote
the point-to-point distance (in meters) yielded by Vincenty's formula
\cite{Vincenty} as $d_V(\cdot,\cdot)$. Then, for any pair of jams, say,
indexed $i$ and $j$, we define the distance between them as
\[
\min \{ d_V(\bz_1, \bz_2);\ \forall \bz_1\in \text{loc}[i],\, \bz_2 \in
\text{loc}[j] \}.  
\] 
Using the distance defined above and setting the quantization level for
``location'' as $3$, we apply the commonly used $K$-means clustering
method \cite{kmeans}, thus obtaining 3 cluster centers as depicted in
Fig. \ref{fig:locClus} (note that, by \eqref{locS} each cluster center
is represented by $4$ red markers).

\begin{table*}[hbt]
	\centering
	\caption{Key features of the detected abnormal jams.} \label{tab1}
	\resizebox{0.9\linewidth}{!}
	{\begin{tabular}{lllllccl}
			\toprule[1.2pt]
			\textbf{index}  & \textbf{start time} & \textbf{detected time} & \textbf{latitude}  & \textbf{longitude} & \textbf{delay (in seconds)} & \textbf{length (in meters)} & \textbf{alert type}   \\
			\midrule[1.2pt]
			788  & 12:25:0.302  & 12:30:0.0 & 42.361951 & -71.117963  & 232.0  & \textcolor{red}{\textbf{3568.0}} & heavy traffic\\
			1502  & 15:35:0.072  & 15:40:0.0 & 42.356275 & -71.119852  & \textcolor{red}{\textbf{585.0}}  & 844.0 & heavy traffic\\
			2412  & 19:25:0.365  & 19:30:0.0 & 42.342549 & -71.085011 & \textcolor{red}{\textbf{643.0}} & \textcolor{magenta}{\textbf{3568.0}} & heavy traffic\\
			3005  & 21:25:0.238  & 21:30:0.0 & 42.349125 &  -71.10778   & 168.0  & \textcolor{red}{\textbf{1962.0}} &  weather hazard\\
			3094  & 21:35:0.267  & 21:40:0.0 & 42.373336 & -71.097731  & \textcolor{red}{\textbf{509.0}}  & 897.0 & road closed \\
			3126  & 21:35:0.326  & 21:40:0.0 &  42.355048 & -71.110335   & \textcolor{red}{\textbf{528.0}}  & 1293.0 & heavy traffic\\
			\bottomrule[1.2pt]
		\end{tabular}}
	\end{table*}

\begin{figure}[htp]  
	\centering
	\begin{subfigure}[b]{0.485\textwidth}
		\includegraphics[width=\textwidth]{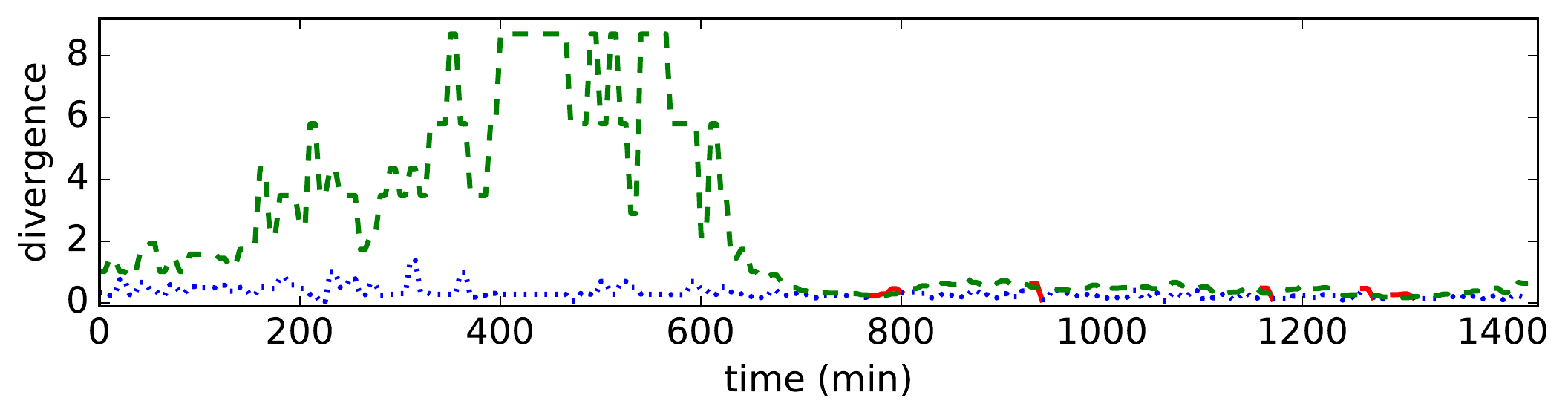}
		\caption{delay}
		\label{wazeDelay}
	\end{subfigure} 
	\begin{subfigure}[b]{0.485\textwidth}
		\includegraphics[width=\textwidth]{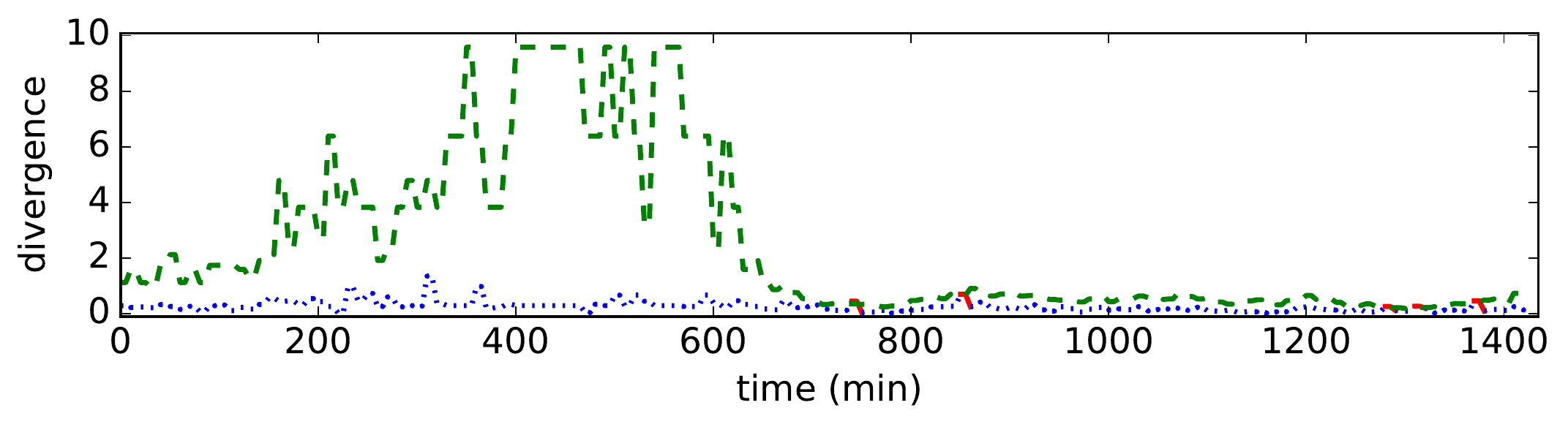}
		\caption{length}
		\label{wazeLength}
	\end{subfigure}   
	\caption{Initial detection results for Waze jams.}
	\label{fig:anoWaze}
\end{figure}

In all our experiments, we take the quantization level to be 1 for
``speed,'' and set the target false alarm rate as $\beta = 0.001$. The
window size is taken as $w_{\text{s}} = 10$ minutes, and the distance
between two consecutive windows is $w_{\text{d}} = 5$ minutes.  To
estimate the PLs, we divide a whole day into 4 subintervals: 5:00-10:00
(AM), 10:00-15:00 (MD), 15:00-19:00 (PM), and 19:00-5:00 (NT).  So, for
each scenario we end up with $4$ PLs, corresponding to the AM peak, the
middle day, the PM peak, and the night, respectively. To calculate the
threshold $\eta_{n,\beta}^{\text{wc}}$ for each detection window, we use
Alg. \ref{alg:thresrob1}.

\subsubsection{Detection Results}
First, let the quantization level for ``delay'' be 2 and for ``length''
be 1. The original sample path has $N = 3 \times 1 \times 2 \times 1 =
6$ states. Thus, we have $N^2 = 36$ states in the transformed
chain. {We use relatively sparse quantization levels for ``delay''
  and ``length'' to avoid unnecessary computational overhead in the
  quantization subroutine for the jam location data.} After running our
algorithm in the initial step, 910 out of 3882 jams are reported within
abnormal detection windows, which correspond to the red segments in
Fig. \ref{wazeDelay}. We then perform a refinement procedure by
selecting jams in these windows with non-typical individual features as
follows. For each selected feature, we calculate the sample mean $\mu$
and sample standard deviation $\sigma$ using the reference data. We then
label as anomalous any jam with feature value exceeding $\mu + 3
\sigma$. We first consider the delay feature. Using the $3\sigma$-rule
on delay, we obtain an anomaly list $\scrL_1$ containing $4$ jams.

Second, let the quantization level for ``delay'' be 1 and for ``length''
be 2. Then, again, the original sample path has $N = 3 \times 1 \times 1
\times 2 = 6$ states, and we have $N^2 = 36$ states in the transformed
chain. After rerunning the algorithm in the initial step, 590 out of
3882 jams are reported within abnormal detection windows, which
correspond to the red segments in Fig. \ref{wazeLength}, and, after
refining by use of the $3\sigma$-rule on the feature ``length'', we end
up with an anomaly list $\scrL_2$ containing $2$ jams.

Finally, we take $\scrL = \scrL_1 \cup \scrL_2$ as our ultimate anomaly
list, which contains $6$ jams in total. By checking the time stamps and
the alarm instances, we see that all of these 6 jams would be reported
as abnormal by our method within $5$ minutes from their start time; this
is satisfactory in a real-time traffic jam anomaly detection
application. Note that we can tune $w_{\text{d}}$ and $w_{\text{s}}$ such that the
detection becomes even faster while maintaining good accuracy in
identifying anomalies. Specifically, smaller $w_{\text{d}}$ leads to faster
detection while $w_{\text{s}}$ should be reasonably big (the number of jams in a
window should at least be comparable to $N^2$). By comparing the
locations and time stamps, we map the jams in the final anomaly list to
the alert data $\scrJ_3$, and find that one of them was reported by
Wazers as ``road closed,'' another as ``weather hazard,'' and all the
others as ``jam heavy traffic.''  In addition, all of them occurred
during non-peak hours. We list the key features of these abnormal jams
in Tab. \ref{tab1}, where the atypical values of the features ``delay''
and ``length'' have been highlighted in bold red. It is worth pointing out
that jam 2412 is reported as abnormal based on ``delay,'' but its length
(highlighted in bold magenta) is also above the threshold for refining the
detection results based on ``length.'' Note also that the latitude and
longitude in each row of Tab. \ref{tab1} represent the closest location
of the Wazer who reported the alert for the corresponding jam (extracted
from the alert data $\scrJ_3$); the shapes of the actual jams have been
visualized as colored bold curves in Fig. \ref{fig:locClus}. While in
this application we do not have ground truth, it is reassuring that the
jams we identify as anomalous have indeed been reported as non-typical
by Wazers. Clearly, depending on how such a detection scheme will be
used by a City's transportation department, our approach provides
flexibility in setting thresholds to adjust the volume of reported
anomalous jams. This volume will largely depend on the resources that
City personnel have to further investigate anomalous jams (e.g., using
cameras) and intervene.

\begin{rem} \label{reAnoWaze} \emph{If we directly apply the $3
    \sigma$-rule on the whole test data without implementing the
    Hoeffding test to obtain a potential anomaly list first, then we
    would very likely end up with too many anomalies, which might
    include undesirable false alarms. Indeed, when we apply the $3
    \sigma$-rule on the whole test data for ``delay'' (resp.,
    ``length''), we obtain 38 (resp., 62) ``anomalies,'' which are
    much more than those in our final anomaly list (6
    only). Thus, including the well-validated Hoeffding test in our
    method ensures a good control of false alarms.}
\end{rem}

\section{Conclusions and Future Work} \label{sec:con}

We have established weak convergence results for the relative entropy in
the Hoeffding test under Markovian assumptions, which enables us to
obtain a {tighter} estimator (compared to the existing estimator based on
Sanov's theorem) for the threshold needed by the test. We have
demonstrated good performance of our estimator by applying the Hoeffding
test in extensive numerical experiments for the purpose of statistical
anomaly detection. The application scenarios involve not only simulated
communication networks, but also real transportation networks. Our work
contributes to enhancing cyber security and helping build smarter
cities.

As for future work, it is of interest to establish theoretical comparison results concerning the tightness of the threshold estimators. The challenge in this direction arises from associating the finite sample-size setting with the asymptotic properties of the Central Limit Theorem and the large deviations results (Sanov's theorem). It is also of interest to conduct rigorous analysis  relating the computation time of the proposed estimation approach to its accuracy. Also, it is possible to consider additional applications.

\appendices
\numberwithin{equation}{section}

\section{Proof of Lemma \ref{le1}} \label{sec:lem1}

Expanding the first $N$ entries of ${\bpi} \bP = {\bpi}$, we obtain
$q_{1i}\sum_{t = 1}^N {{\pi _{t1}}} = {\pi _{1i}}$, $i = 1,
\ldots, N$.  Summing up both sides of these equations, it follows
\begin{equation}
 \Big( \tsum_{i = 1}^N q_{1i}\Big) \Big( \tsum_{t = 1}^N \pi _{t1}\Big)
 = \tsum_{t = 1}^N \pi _{1t}.  \label{3.2} 
\end{equation}
Noticing $\sum_{i = 1}^N q_{1i} = 1$, \eqref{3.2} implies $\sum_{t =
  1}^N {{\pi _{t1}}} = \sum_{t = 1}^N {{\pi _{1t}}}$, which, together
with $q_{11}\sum_{t = 1}^N {{\pi _{t1}}} = {\pi _{11}}$, yields
\[
\frac{{{\pi _{11}}}}{{\tsum_{t = 1}^N {{\pi _{1t}}} }} = \frac{{{\pi
      _{11}}}}{{\tsum_{t = 1}^N {{\pi _{t1}}} }} = {q_{11}}.
\]
Similarly, we can show \eqref{3} holds for all the other $(i,\,j)$'s.

\section{Proof of Lemma \ref{le2}}  \label{sec:lem2}

This can be established by applying \cite[Corollary
1]{jones2004markov}. Noting $f_k(\cdot)$ is an indicator function, thus
Borel measurable and bounded, and the chain $\bZ$ is uniformly ergodic,
we see that, $\exists B \in ( {0,\infty })$ s.t. $|
  {{f_k}( Z )} | \leq B, \, \forall Z$, implying that
$\mathbbm{E}[ {{{| {{f_k}( Z)}|}^3}} ] \leq {B^3} < \infty $, and
\cite[(3)]{jones2004markov} holds 
with $M(\cdot)$ bounded, leading to $\mathbbm{E}[ M ] <
\infty $, and $\gamma(n) = t^n$ for some $t \in (0, 1)$, indicating that
$\sum\nolimits_n {{{( {\gamma ( n )} )}^{{1
        \mathord{/ {\vphantom {1 3}} .
          \kern-\nulldelimiterspace} 3}}}} = \sum\nolimits_n {{t^{{n
        \mathord{/ {\vphantom {n 3}} .
          \kern-\nulldelimiterspace} 3}}}} = \sum\nolimits_n {{{(
        {{t^{{1 \mathord{/ {\vphantom {1 3}} .
                  \kern-\nulldelimiterspace} 3}}}} )}^n}} < \infty
$. Thus, all the conditions needed by \cite[Corollary
1]{jones2004markov} are satisfied.

\section{Proof of Lemma \ref{th1}}  \label{sec:thm1}

We can directly extend Lemma \ref{le2} to the multidimensional case (see
\cite[Chap. 8]{GeyerCourseNotes}). In particular, under
Assumption \ref{ass1}, \eqref{4} holds with $\bLambda$ given by
\begin{equation} 
\bLambda =  {\bLambda ^{\scriptscriptstyle{(0)}}} + \tsum_{m =
  1}^\infty  {{\bLambda ^{\scriptscriptstyle{(m)}}}}, 
\label{40}
\end{equation}
where ${\bLambda ^{\scriptscriptstyle{(0)}}}$ and ${\bLambda
  ^{\scriptscriptstyle{(m)}}}$ are specified, respectively,
by 
\begin{align*}
{\bLambda ^{\scriptscriptstyle{(0)}}} = & [ {\operatorname{Cov} (
  {{f_i}( {{Z_1}} ),~{f_j}( {{Z_1}} )} )} ]_{i,\,j = 1}^{{N^2}},\\
{\bLambda ^{\scriptscriptstyle{(m)}}} = &  [{\mathop{\rm Cov}\nolimits} ( {{f_i}( {{Z_1}} ),{f_j}( {{Z_{1 + m}}} )} ) \notag\\
 & \phantom{[ } 
+ {\mathop{\rm Cov}\nolimits} ( {{f_j}( {{Z_1}} ),{f_i}( {{Z_{1 + m}}}
  )} )]_{i,\,j = 1}^{{N^2}},
 \quad m = 1,2, \ldots . \notag
\end{align*}

Let the subscript $ij$ denote the $(i,j)$ elements of the matrices
$\bLambda, \bLambda^{\scriptscriptstyle{(0)}},
\bLambda^{\scriptscriptstyle{(m)}}$. By the Markovian properties, after some direct algebra, 
for $i,\,j = 1, \ldots ,{N^2}$ we obtain $\Lambda
^{\scriptscriptstyle{(0)}}_{ij} = {{\tilde \pi }_i}( {{{\bI}_{ij}} -
  {{\tilde \pi }_j}} )$ and
\[
\Lambda ^{\scriptscriptstyle{(m)}}_{ij} = {{\tilde \pi }_i}(
{{\bP}_{ij}^m - {{\tilde \pi }_j}} ) + {{\tilde \pi }_j} ( {{\bP}_{ji}^m
  - {{\tilde \pi }_i}} ),\quad m = 1,2, \ldots. 
\]

\section*{Acknowledgments}

We thank Jing Wang for his contributions and help in developing the
software package SADIT \cite{SADIT}. We also thank the DoIT of the City
of Boston, Chris Osgood, Alex Chen, and Connor McKay for supplying the
Waze data. We thank Athanasios Tsiligkaridis for his help in deriving
Tab.~\ref{tab1} in Sec.~\ref{sec:anoWaze}. We finally thank the
anonymous reviewers for useful comments on preliminary versions of this
paper.

\bibliographystyle{IEEEtran}


\bibliography{bib}

\end{document}